# AIRCC-Clim: a user-friendly tool for generating regional probabilistic climate change scenarios and risk measures

*Francisco Estrada[1,2,3]\*, Oscar Calderón-Bustamante[1], Wouter Botzen[2,4], Julián A. Velasco[1], Richard S.J. Tol[5,6,7,8,9]*

[1]Centro de Ciencias de la Atmosfera, Universidad Nacional Autónoma de México, CDMX, Mexico; [2]Institute for Environmental Studies, VU Amsterdam, Amsterdam, the Netherlands; [3]Programa de Investigación en Cambio Climático, Universidad Nacional Autónoma de México, CDMX, Mexico; [4]Utrecht University School of Economics (U.S.E.), Utrecht University, Utrecht, Netherlands; [5]Department of Economics, University of Sussex, Falmer, UK; [6]Department of Spatial Economics, Vrije Universiteit, Amsterdam, The Netherlands; [7]Tinbergen Institute, Amsterdam, The Netherlands; [8]CESifo, Munich, Germany; [9]Payne Institute for Public Policy, Colorado School of Mines, Golden, CO, USA

*Corresponding autor: feporrua@atmosfera.unam.mx

Highlights

- A new tool for generating probabilistic regional climate change scenarios and risk measures
- AIRCC-Clim is a standalone, user-friendly software designed for a variety of applications including impact, vulnerability and adaptation assessments, integrated assessment modelling,
- It includes a graphical interface that allows the quick evaluation of the consequences on global and regional climate of user-defined experiments of international mitigation policies.

Abstract

Complex physical models are the most advanced tools available for producing realistic simulations of the climate system. However, such levels of realism imply high computational cost and restrictions on their use for policymaking and risk assessment. Two central characteristics of climate change are uncertainty and that it is a dynamic problem in which international actions can significantly alter climate projections and information needs, including partial and full compliance of global climate goals. Here we present AIRCC-Clim, a simple climate model emulator that produces regional probabilistic climate change projections of monthly and annual temperature and precipitation, as well as risk measures, based both on standard and user-defined emissions scenarios for six greenhouse gases. AIRCC-Clim emulates 37 atmosphere-ocean coupled general circulation models with low computational and technical requirements for the user. This standalone, user-friendly software is designed for a variety of applications including impact assessments, climate policy evaluation and integrated assessment modelling.

Software availability: AIRCC-Clim can be downloaded at no cost from https://sites.google.com/view/aircc-lab-airccclim/aircc-clim

1. Introduction

Climate change projections are one of the main inputs for assessing the potential consequences of different socioeconomic development pathways and international climate policy on natural and human systems. Due to the complexity of the systems involved and their interactions, climate projections are inherently uncertain (Curry and Webster, 2011; Gay and Estrada, 2010a; Knutti and Sedláček, 2012). Moreover, computational and technical costs of state-of-the-art physical models allow exploration of only a small fraction of the range of possible climate futures and hinder assessing risk through probabilistic scenarios (Knutti et al., 2010; Sanderson et al., 2015). For most decision-makers and researchers, these costs make it infeasible to explore how current and hypothetical changes in international mitigation policy can influence future warming and its consequences for society. In a time of proactive international mitigation policy, the dynamic nature of projecting future climate becomes even more evident and decision-making requirements can go beyond fixed illustrative emissions scenarios, such as the RCPs (Estrada and Botzen, 2021). For example, Nationally Determined Contributions (NDCs) that represent greenhouse gas emission reductions that countries promise as their contributions to the Paris Agreement are currently a key focus of international climate policy. Moreover, due to the nonlinearity of most climate impacts, even small deviations from a high-warming trajectory can produce large changes in the associated impacts (Estrada and Botzen, 2021; Ignjacevic et al., 2021).

International efforts such as the Coupled Model Intercomparison Project (CMIP), which build and host large databases of climate models' simulations publicly available, have significantly contributed to improving the accessibility and utilization of climate scenarios for the wider research community and decision-makers (Knutti and Sedláček, 2012; Thomas F. Stocker et al., 2013; Taylor et al., 2012). However, many users still face the problem of processing large datasets for adapting them to their particular needs (e.g., temporal frequency and spatial domains). As such, studies and decisions are commonly based on a few illustrative greenhouse gases emissions trajectories and a handful of climate models' simulations which are often selected due to their availability and ease of use, such as WorldClim (Fick and Hijmans, 2017). Such a selection of climate model runs can hardly provide a good representation of uncertainty and indicate if a model's projections for a given region may represent extreme realizations in comparison to the majority of other models (Sanderson et al., 2015; Weigel et al., 2010). Even in cases when climate models' performance is assessed, the resulting selection of models does not guaranty that those projections of future climate are reliable (Altamirano del Carmen et al., 2021). Climate model selection remains an unresolved problem as commonly used metrics can be non-informative about the models' ability to reproduce the observed climate change signal and indicate much less about their ability for projecting future climate (Knutti et al., 2010).

Uncertainty is a key characteristic of climate change and how it is understood and included in climate impact assessments can have profound effects on the estimates of the consequences of this phenomenon and on the design of policies to address these consequences (Gay and Estrada, 2010a; IPCC-TGICA, 2007). The development of tools and methods for better uncertainty management and for improving the usefulness of the large (thousands of terabytes) databases that are currently available are increasingly relevant research topics. More efficient, simple, and flexible approaches for taking advantage of the available databases can transform data into useful information and knowledge for better assessments of impacts, risks, and climate policy options. Reduced complexity models and emulators of more advanced models allow exploring –at low computational and technical costs for the user– a wide range of possible futures and emissions trajectories, parameterizations, as well as probabilistic assessment of risks for natural and human systems (Blanc, 2017; Estrada et al., 2020; Meinshausen et al., 2011a). Integrated assessment modelling benefits from such models and emulators to provide tools for supporting decision-making and providing estimates for cases in which complex climate/impacts model runs are not available.

A notable example of the usefulness of such models is the MAGICC software which has made significant contributions to climate change research, particularly in impact, vulnerability and adaptation (IVA) assessments, and integrated assessment modelling (Meinshausen et al., 2011a, 2011b; Wigley, 1995). The MAGICC software has been regularly used in the IPCC reports when simulations produced by general circulation climate models are not available and also in the national climate change assessments of several countries (Conde et al., 2011; IPCC, 2021, 2018).

Here we present AIRCC-Clim (Assessment of Impacts and Risks of Climate Change – Probabilistic Climate Model Emulator), a simple and flexible climate model emulator for producing probabilistic regional projections of monthly and annual temperature and precipitation, as well as risk measures. AIRCC-Clim emulates the results from 37 atmosphere-ocean coupled general circulation models included in the Fifth Assessment Report of the IPCC (Thomas F. Stocker et al., 2013) with low computational and technical requirements on the user. It is a standalone software for Windows and Linux operating systems that requires no programming or advanced technical skills from the user and runs on computers with standard memory and processing resources. AIRCC-Clim is designed for a variety of applications including IVA assessments, integrated assessment modelling, and the quick evaluation of the consequences on global and regional climate of user-defined experiments of international mitigation.

The remainder of this paper is structured as follows: Section 2 describes the AIRCC-Clim model structure and describes in detail each of its modules in terms of data sources, modelling approaches and methods, and their output. This section also describes the file options for exporting climate projections and their characteristics. An application of the model is shown in Section 3 in which the benefits of stringent international mitigation efforts are illustrated in terms of avoided warming, changes in precipitation and risk reduction. Section 4 concludes and discusses model extensions and integration with IVA models.

2. Model structure, data and methods

AIRCC-Clim is composed of four main modules: greenhouse gas emission scenario editor; global climate models, a regional scenario generator; and a climate risk index generator (Figure 1). In the following paragraphs each module is discussed, including detailed descriptions of the modelling approaches, methods and data sources.

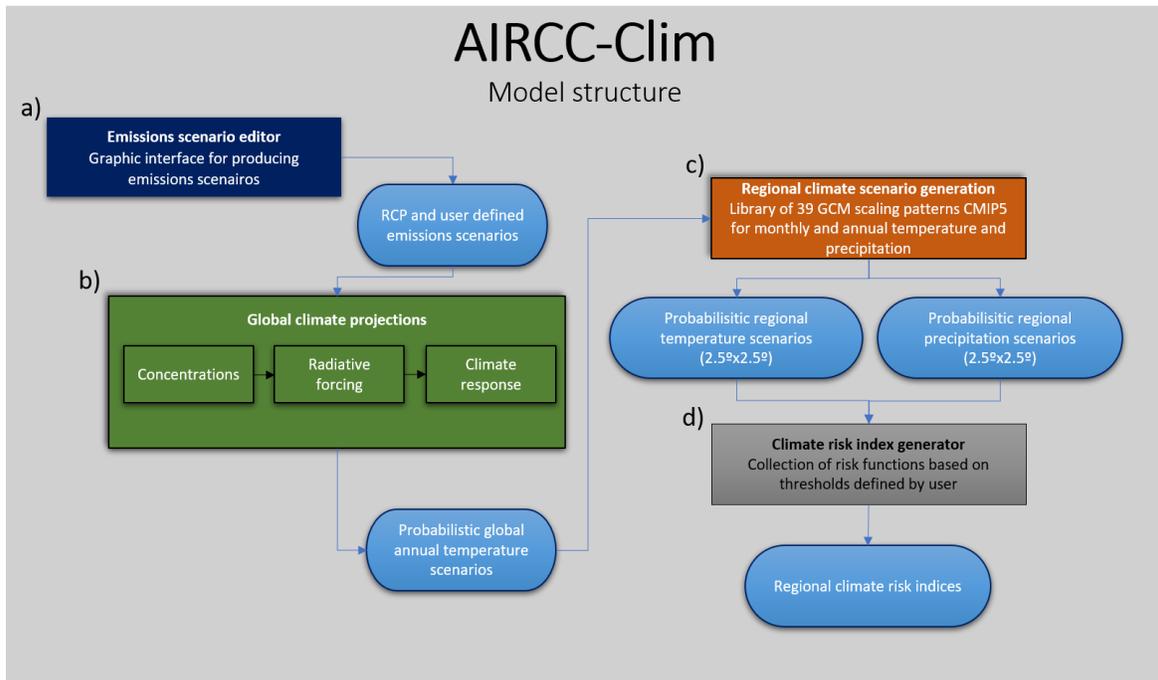

Figure 1. Schematic representation of AIRCC-Clim model structure. Calculation modules are represented by rectangles while output is denoted by rectangles with rounded corners. The four modules are: a) the emissions scenario editor; b) the climate module which produces annual global temperature change estimates; c) the regional climate scenario generator which produces probabilistic monthly and annual climate projections, and; d) the climate risk index generator that calculates risk measures defined by the user.

2.1. Emissions scenario editor

AIRCC-Clim offers a graphic interface for constructing user-defined emissions scenarios for six climate changing substances: Carbon dioxide ($CO_2$ in MtC/yr), methane ($CH_4$ in $MtCH_4$/yr), nitrous oxide ($N_2O$; $MtN_2O$-N/yr), chlorofluorocarbons ($CFC_{11}$, $CFC_{12}$ in kt/yr), and sulfur hexafluoride ($SF_6$ in kt/yr). Four RCP emissions trajectories are included by default and the user can select one of them as starting point for editing. As shown in Figure 2 AIRCC-Clim has three input options for emissions: 1) by means of editing the values displayed on a table; 2) by selecting and plotting the substance of interest and modifying the emissions trajectories directly in a graph; 3) loading an Excel file (.xlsx) with user-defined emissions trajectories for each gas in a predefined format (see the AIRCC-Clim user guide included in the SI). The modified emissions scenario is saved in an Excel (.xlsx) file and used for generating the corresponding climate projections in AIRCC-Clim.

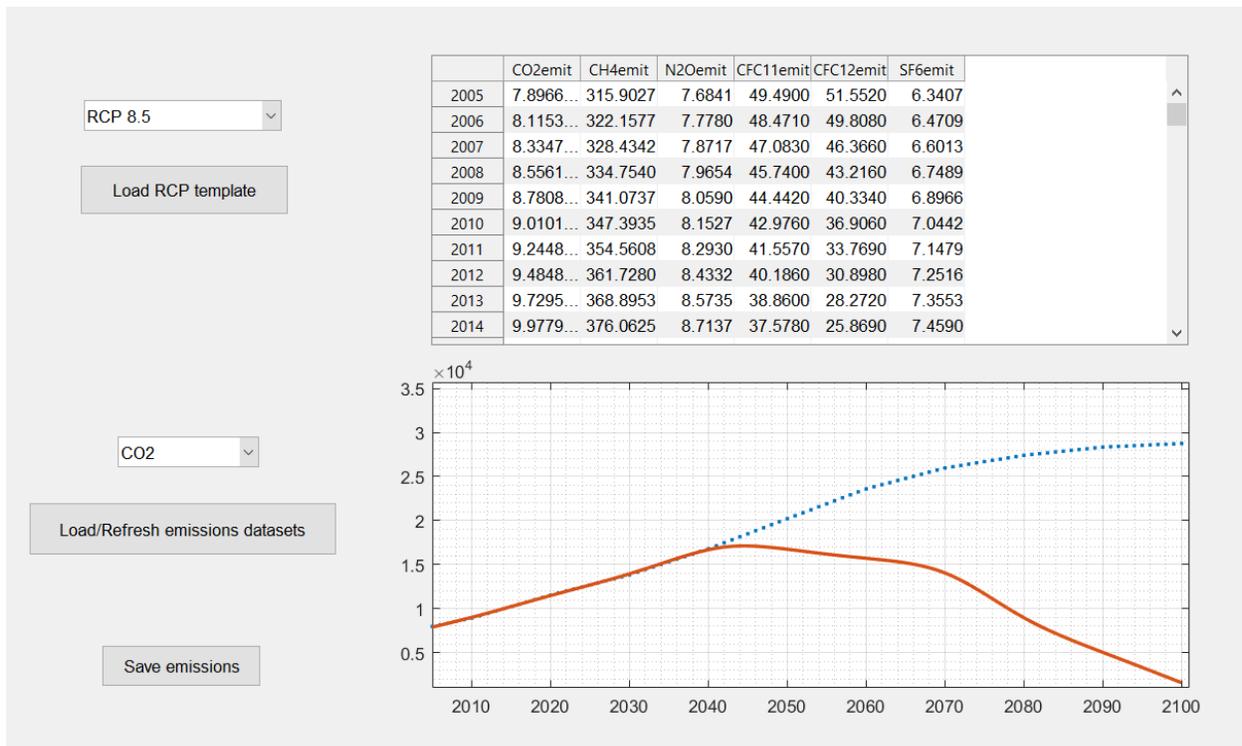

Figure 2. AIRCC-Clim's graphic emissions editor. Four RCP emissions trajectories can be loaded and edited by the user by means of: 1) modifying the values in the editor's table (upper part); 2) by changing the trajectory in an interactive graph (lower part) and; 3) loading an Excel file with user-defined emissions trajectories. In the lower part of the figure, the dashed blue line shows the default values for the selected RCP scenario, while the continuous red line shows the trajectory edited by the user.

## 2.2. Global climate projections

Global temperature projections in AIRCC-Clim can be generated by running a simple climate model, as well as by using precalculated projections from two reduced complexity climate models. As described in the following paragraphs, in all cases probabilistic projections are constructed by means of stochastic simulation to represent uncertainty in the climate sensitivity (CS) parameter.

### 2.2.1. Modified Schneider-Thompson model

The Schneider-Thompson (ST) simple climate model (Schneider and Thompson, 1981) includes three components that allows to calculate the atmospheric concentrations of greenhouse gases, the corresponding change in radiative forcing and the resulting increase in global temperature. Due to its flexibility and low computational cost, this climate model (and modified versions of it) is used in some of the most popular integrated assessment models, such as FUND and RICE (Anthoff and Tol, 2014; Nordhaus and Boyer, 2003).

The version of the ST model in AIRCC-Clim builds upon that used in the FUND integrated assessment model (Tol and Fankhauser, 1998) which is available at http://www.fund-model.org/. In this section, the ST model and the modifications that are introduced in AIRCC-Clim.

*Calculation of atmospheric concentrations and radiative forcing*

A five box-model (Tol, 2019) is used to convert annual emissions of $CO_2$ (MtC) into atmospheric concentrations (ppm). The initial value for $CO_2$ concentrations is 278 ppm which represent preindustrial times (circa 1750). The carbon cycle model is represented by the following equation:

$$C_{i,t}^{CO_2} = (1 - \alpha_i)C_{i,t-1}^{CO_2} + \gamma_i \beta E_t^{CO_2} \quad (1)$$

$$C_t^{CO_2} = \sum_{i=1}^{5} C_{i,t}^{CO_2} \quad (2)$$

where $C_{i,t}^{CO_2}$ represents the atmospheric concentrations of $CO_2$ in box $i=1,...,5$ at time t, $E_t$ are the $CO_2$ emissions at time t, $\alpha_i$ determines how long $CO_2$ remain in box $i$, while $\gamma_i$ is the proportion of emissions that enter box $i$, and $\beta$ is a scale parameter ($\beta = 0.00045$). These boxes are characterized by different decay times that resemble the slow and fast components of the carbon cycle. However, these boxes do not represent physical processes and are just simple mathematical devices that allow to approximate the observed concentrations (Schneider and Thompson, 1981; Tol, 2019). $C_t^{CO_2}$ is the total concentration of $CO_2$ in the atmosphere at time *t*. Parameter values are reproduced in Table S1. The radiative forcing from CO2 is calculated as:

$$F_t^{CO_2} = 5.35 \ln\left(\frac{C_{i,t}^{CO_2}}{C_{pre}^{CO_2}}\right) \quad (3)$$

Where $C_{pre}^{CO_2} = 278$ represents the preindustrial atmospheric concentrations of $CO_2$.

CH4 concentrations are calculated using the following equation:

$$C_t^{CH_4} = (1 - a_1)C_{t-1}^{CH_4} + a_1 C_{pre}^{CH_4} + b_1 E_t^{CH_4} \quad (4)$$

Where $C_t^{CH_4}$ are the atmospheric concentrations of CH4 at time t, $C_{pre}^{CH_4} = 721.89$ is the preindustrial concentrations of CH4, $E_t^{CH_4}$ are the emissions of CH4 at time *t*, $a_1 = 0.1057$ represents the permanence of CH4 in the atmosphere and $b_1 = 0.3597$ is a scaling factor.

The atmospheric concentrations of N2O are calculated using an equation similar to that of CH4:

$$C_t^{N_2O} = (1 - a_2)C_{t-1}^{N_2O} + a_2 C_{pre}^{N_2O} + b_2 E_t^{N_2O} \quad (5)$$

Where the emissions and atmospheric concentrations of N2O are denoted by $E_t^{N_2O}$ and $C_t^{N_2O}$, respectively, while $C_{pre}^{N_2O} = 272.96$, $a_2 = 1/120$ and $b_2 = 0.1550$.

Calculating the radiative forcing of CH4 and N2O involves an interaction term between these gases to account for their overlap in the absorption bands as represented in the following equations:

$$Int_t^{CH_4} = f(M, N_0) - f(M_0, N_0) \quad (6)$$

$$Int_t^{N_2O} = f(M_0, N) - f(M_0, N_0)' \qquad (7)$$

Where these interaction functions are given by:

$$f(M, N_0) = p_1 ln\left[1 + p_2\left(C_t^{CH_4} C_{t=0}^{N_2O}\right)^{0.75} + p_3 C_t^{CH_4}\left(C_t^{CH_4} C_{t=0}^{N_2O}\right)^{1.52}\right] \qquad (8)$$

$$f(M_0, N) = p_1 ln\left[1 + p_2\left(C_{t=0}^{CH_4} C_t^{N_2O}\right)^{0.75} + p_3 C_t^{CH_4}\left(C_{t=0}^{CH_4} C_t^{N_2O}\right)^{1.52}\right] \qquad (9)$$

$$f(M_0, N_0) = p_1 ln\left[1 + p_2\left(C_{pre}^{CH_4} C_{t=0}^{N_2O}\right)^{0.75} + p_3 C_{pre}^{CH_4}\left(C_{pre}^{CH_4} C_{t=0}^{N_2O}\right)^{1.52}\right] \qquad (10)$$

$$f(M_0, N_0)' = p_1 ln\left[1 + p_2\left(C_{t=0}^{CH_4} C_{pre}^{N_2O}\right)^{0.75} + p_3 C_{pre}^{N_2O}\left(C_{t=0}^{CH_4} C_{pre}^{N_2O}\right)^{1.52}\right] \qquad (11)$$

With $p_1 = 0.47$, $p_2 = 2.01 * 10^{-5}$, $p_3 = 5.31 * 10^{-15}$. The radiative forcing of CH4 and N20 is calculated as:

$$F_t^{CH_4} = 0.036\left[\left(C_t^{CH_4}\right)^{0.5} - \left(C_{pre}^{CH_4}\right)^{0.5}\right] - Int_t^{CH_4} \qquad (12)$$

$$F_t^{N_2O} = 0.12\left[\left(C_t^{N_2O}\right)^{0.5} - \left(C_{pre}^{N_2O}\right)^{0.5}\right] - Int_t^{N_2O} \qquad (13)$$

CFC$_{11}$ and CFC$_{12}$ concentrations are calculated as follows:

$$C_t^{CFC_{11}} = (1 - a_3)C_{t-1}^{CFC_{11}} + b_3 E_t^{CFC_{11}} \qquad (14)$$

$$C_t^{CFC_{12}} = (1 - a_4)C_{t-1}^{CFC_{12}} + b_4 E_t^{CFC_{12}} \qquad (15)$$

In which $a_3$ and $a_4$ are equal to $1/45$ and $1/100$ while $b_3 = 0.0423$ and $b_4 = 0.0481$. The radiative forcing from CFCs is calculated by multiplying it by a scaling factor equal to $0.25/1000$ in the case of CFC$_{11}$ and $0.32/1000$ for CFC$_{12}$.

Finally, the concentrations of $SF_6$ are obtained using the following equation:

$$C_t^{SF_6} = (1 - a_5)C_{t-1}^{SF_6} + a_5 C_{pre}^{SF_6} + b_5 E_t^{SF_6} \qquad (16)$$

Where $a_5 = 1/3200$ and $b_5 = 0.0393$. The radiative forcing of $SF_6$ is calculated as $F_t^{SF_6} = 0.00052 * C_t^{SF_6}$.

*Calculation of global temperature change*

The ST model uses the following two interlinked equations to produce annual mean global air and ocean surface temperatures:

$$T_t^A = T_{t-1}^A + \lambda_1(\lambda_2 F_t - T_{t-1}^A) + \lambda_3(T_{t-1}^O - T_{t-1}^A) \qquad (17)$$

$$T_t^O = T_{t-1}^O + \lambda_4(T_{t-1}^A - T_{t-1}^O) \qquad (18)$$

Following Tol (2019) we use the following parameter values as the initial calibration: $\lambda_1 = 0.0256$, $\lambda_2 = 1.14891$, $\lambda_3 = 0.00738$, and $\lambda_4 = 0.00568$. The CS of this model is calculated as $\lambda_2[5.35\ln(2)]$ and it is the main parameter for calibrating the model to reproduce observed or projected annual mean global air surface temperature. A limitation with this approach is that the accuracy of its projections will depend on which temperature series is used for calibration. To illustrate this, we use the projections obtained from the MAGICC6 model for the RCP8.5, RCP6, RCP4.5 and RCP2.6 for the period 2005-2100. The calibration procedure used was to minimize the sum of the squared residuals between the TS and the MAGICC6 projections using ordinary least squares. The CS values that calculated for each scenario are 3.27 (RCP8.5), 2.70 (RCP6), 2.64 (RCP4.5) and 2.23 (RCP2.6). Choosing any single value of this parameter to project all the RCP scenarios would lead to over- or underestimating future temperatures, particularly in the case of extreme scenarios (RCP2.6 and RCP8.5). As such, this approach could have an impact on, for example, the evaluation of the benefits of mitigation policies.

The level of warming is closely related to the total cumulative emissions of CO2 (TCRE). The relationship between global temperature change and TCRE is relatively constant over time and independent from the emissions scenario used, but it is dependent of the global climate model CS and transient climate response (Collins et al., 2013). Figure 1 shows that in the case of MAGICC6 this assumption holds as the cumulative amount of CO2 is linearly related to the estimated CS. Given this relationship, we propose a dynamic CS parameter for the ST model that is calculated based on the cumulative CO2 emissions, which are dominant in any RCP, SRES or other realistic emissions scenario. The CS is thus calculated prior to projecting global temperatures with the ST model. The regression equation that relates the CS and the cumulative CO2 emissions is $CS = 2.05 + 6.18 \times 10^{-7} \sum CO_2$ ($R^2 = 0.96$; Figure 2). To get a better approximation for the lowest and highest cumulative scenarios, we also fitted a line based on the CS values for RCP2.6 and the RCP8.5. The results were compared to the projections in the IPCC Fifth Assessment Report (AR5) and were optimized to reproduce the best estimates. The final extreme CS values are 2.1°C and 3.3°C for the RCP2.6 and RCP8.5 scenarios, respectively. The final equation relating CS and cumulative CO2 emissions is

$$CS = 1.85 + 7.51 \times 10^{-7} \sum CO_2 \qquad (19)$$

In which the CS values are restricted to the interval [2.1, 3.3]. Table 2 shows the projections of the modified ST model and the best estimates and likely ranges in the AR5. For all four RCP emissions scenarios and both short- and long-term horizons, the average difference is about 0.08ºC and the maximum 0.23ºC. This illustrates that the modified ST model very closely resembles the IPCC AR5 temperature projections.

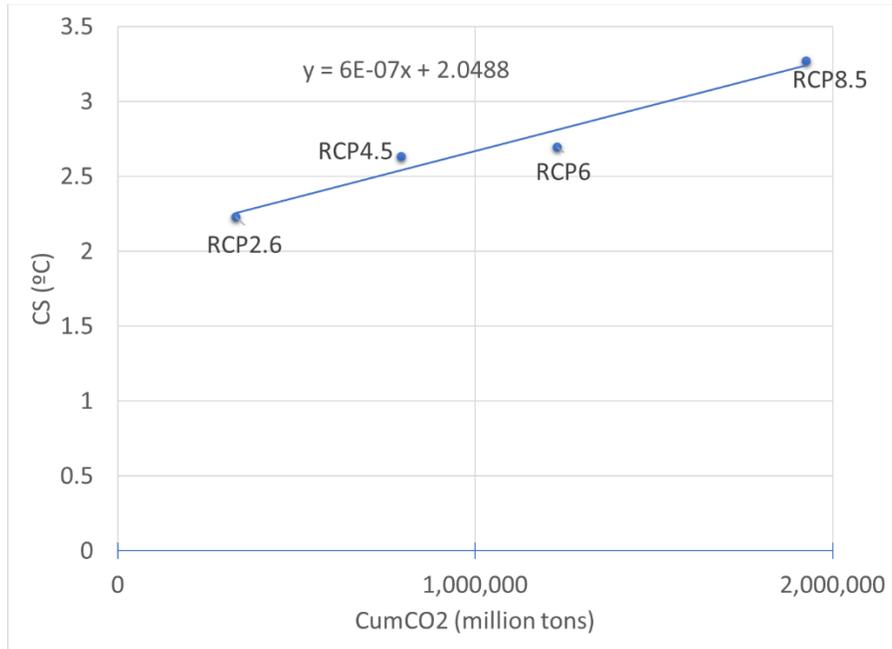

Figure 2. Scatter plot between cumulative CO2 emissions and the climate sensitivity that minimizes the sum of squared errors between the TS and the MAGICC6 projections.

Table 1. Comparison between the best estimates and likely ranges of annual mean global air surface temperature reported in the IPCC's AR5 and those obtained with the modified ST model.

|  |  | 2046-2065 | | 2081-2100 | |
| --- | --- | --- | --- | --- | --- |
|  |  | Best estimate | Likely range | Best estimate | Likely range |
| RCP2.6 | IPCC | 1.61 | 1.01 – 2.21 | 1.61 | 0.91 – 2.31 |
|  | ST* | 1.56 | 1.06 – 2.10 | 1.67 | 1.10 – 2.31 |
| RCP4.5 | IPCC | 2.01 | 1.51 – 2.61 | 2.41 | 1.71 – 3.21 |
|  | ST* | 1.94 | 1.39 – 2.54 | 2.44 | 1.70 – 3.25 |
| RCP6 | IPCC | 1.91 | 1.41 – 2.41 | 2.81 | 2.01 – 3.71 |
|  | ST* | 2.15 | 1.59 – 2.75 | 3.01 | 2.17 – 3.92 |
| RCP85 | IPCC | 2.61 | 2.01 – 3.21 | 4.31 | 3.21 – 5.41 |
|  | ST* | 2.80 | 2.17 – 3.48 | 4.32 | 3.30 – 5.43 |

To extend the ST model to account for the uncertainty in CS values and to produce probabilistic projections, we represent the model's CS with a triangular distribution. In this distribution the most likely value is given by equation 19 above, and the lower and upper limits are the calculated CS value plus/minus some constants that are to be calibrated. The equations for the upper ($CS^{high}$) and lower limits ($CS^{low}$) of the triangular distribution for CS are:

$$CS^{low} = CS + h_1 \qquad (20)$$

$$CS^{high} = CS + h_2 \qquad (21)$$

The $h_1$ and $h_2$ constants are calibrated to match the likely ranges reported in the IPCC's AR5. Simulations of 10,000 realizations were used and the 5$^{th}$ and 95$^{th}$ percentiles were chosen to construct the ST likely ranges. The parameter values that provided good fit for all RCPs are $h_1 = 1.1$ and $h_2 = 1.5$. Table 1 shows that the likely ranges of the modified ST model closely reproduce those reported in the IPCC for both short- and long-term horizons, showing almost exact overlap with differences in upper and lower limits typically smaller than 0.2ºC.

2.2.2. Generating probabilistic global temperature projections using precalculated runs from MAGICC6 and the Thermodynamic Climate Model.

The MAGICC6 and the Thermodynamic Climate Model (TCM) are reduced-complexity climate models that were designed for different objectives. MAGICC is composed of a set of coupled models that include gas cycles and climate and ice-melt models designed to explore the effects of anthropogenic emissions of greenhouse gas concentrations, radiative forcing, and changes in global mean annual temperature and sea level (Meinshausen et al., 2011a, 2011b; Wigley, 1995). The TCM was originally conceived as a weather forecast model for the northern hemisphere and then extended to the global scale for studying the climate of Earth and of other planets in the Solar System (Adem, 1991). In the case of MAGICC6, precalculated projections of global temperature are included in AIRCC-Clim for the emissions scenarios RCP2.6, RCP4.5 RCP6 and RCP8.5, while in the case of the TCM the precalculated that were available are the RCP4.5, RCP6 and RCP8.5.

To account for the uncertainty in CS values and to produce probabilistic projections, we propose a simple method based on linear regression and statistical simulation. Due to the availability of climate models' output we base our calculations on MAGICC6. Projections for each RCP scenario were obtained using of MAGICC6 for three different CS values that represent medium CS (3.0ºC), low CS (1.5ºC) and high CS (4.5ºC). To emulate MAGICC's results for high and low climate sensitivities, we propose the use of some scaling weights $w$ that would approximate them when multiplied by the global temperature obtained using a medium CS value. Such weights can be obtained by means of the following regression:

$$T_t^{sens*} = \omega T_t^{medium} + \varepsilon_t \qquad (22)$$

Where $T_t^{sens*}$ is the global temperature projection obtained with MAGICC6 using either low or high CS, while $T_t^{medium}$ corresponds to the global temperature projection obtained with medium CS; $\omega$ is the slope parameter and $\varepsilon_t$ are the regression residuals. In all cases, the $R^2$ is higher than 0.99; note that the objective of these models is not to make inference about parameter values, but just to produce a close fit of projections using different values of CS. The values of parameter $\omega$ for the different combinations of CS and for each RCP emissions scenario are provided in Table 2. The estimated parameter values are very similar for different RCPs with an average value of 1.34 and 0.57 for high and low CS values, respectively. As shown in Figure 3, these values allow to closely approximate the simulations of MAGICC6 produced using high/low CS values by scaling a simulation of the same model calculated with medium sensitivity. The approximation is less precise when used on scenarios that lead to stabilization (Figure 3d), but the error is still very small (0.17ºC for the high CS projection using the RCP2.6).

Table 2. Estimated slope parameter values of regression $T_t^{sens*} = \omega T_t^{medium} + \varepsilon_t$ for annual mean air surface global temperature projections using MAGICC6.

| Scenario | $\omega$ (CS=4.5) | $\omega$ (CS=1.5) |
| --- | --- | --- |
| RCP8.5 | 1.3369 | 0.5762 |
| RCP6 | 1.3345 | 0.5720 |
| RCP4.5 | 1.3430 | 0.5652 |
| RCP3PD | 1.3510 | 0.5582 |
| Mean | 1.3414 | 0.5679 |

To produce probabilistic scenarios using the ST model, we use a triangular distribution to scale the MAGICC6 runs obtained with a medium CS. The parameters of the triangular distribution are 1 for the mode or most likely value and, 0.5679 and 1.3414 for the lower and upper limits, respectively (Table 2). These parameter values allow to emulate the projections that would be obtained with MAGICC6 randomly choosing CS values contained in the interval 1.5ºC to 4.5ºC, assigning a larger probability of occurrence to values that are closer to the medium sensitivity of 3ºC. CS is high uncertainty (Cox et al., 2018; Freeman et al., 2015; Friedrich et al., 2016; Knutti et al., 2017; Lewis and Curry, 2015; Rogelj et al., 2014; Tan et al., 2016). However, the consensus is that the CS is probably within the 1.5ºC to 4.5ºC interval with a central estimate of 3ºC (Bony et al., 2013; Callendar, 1938; Freeman et al., 2015; Santer et al., 2019; Stocker et al., 2013). This CS range encompasses the range produced by the state-of-the-art models included in the Coupled Model Intercomparison Project 5 (CMIP5) (Jonko et al., 2018; Stocker et al., 2013).

Given that in the TCM the CS is not an explicit parameter as in MAGICC6, but an emerging property of the model, there are no high/low CS projections that can be used to find the corresponding parameters in Table 2. As such we apply the same scaling parameters we found for MAGICC6 to represent uncertainty in CS values and to extend the TCM projections to be probabilistic.

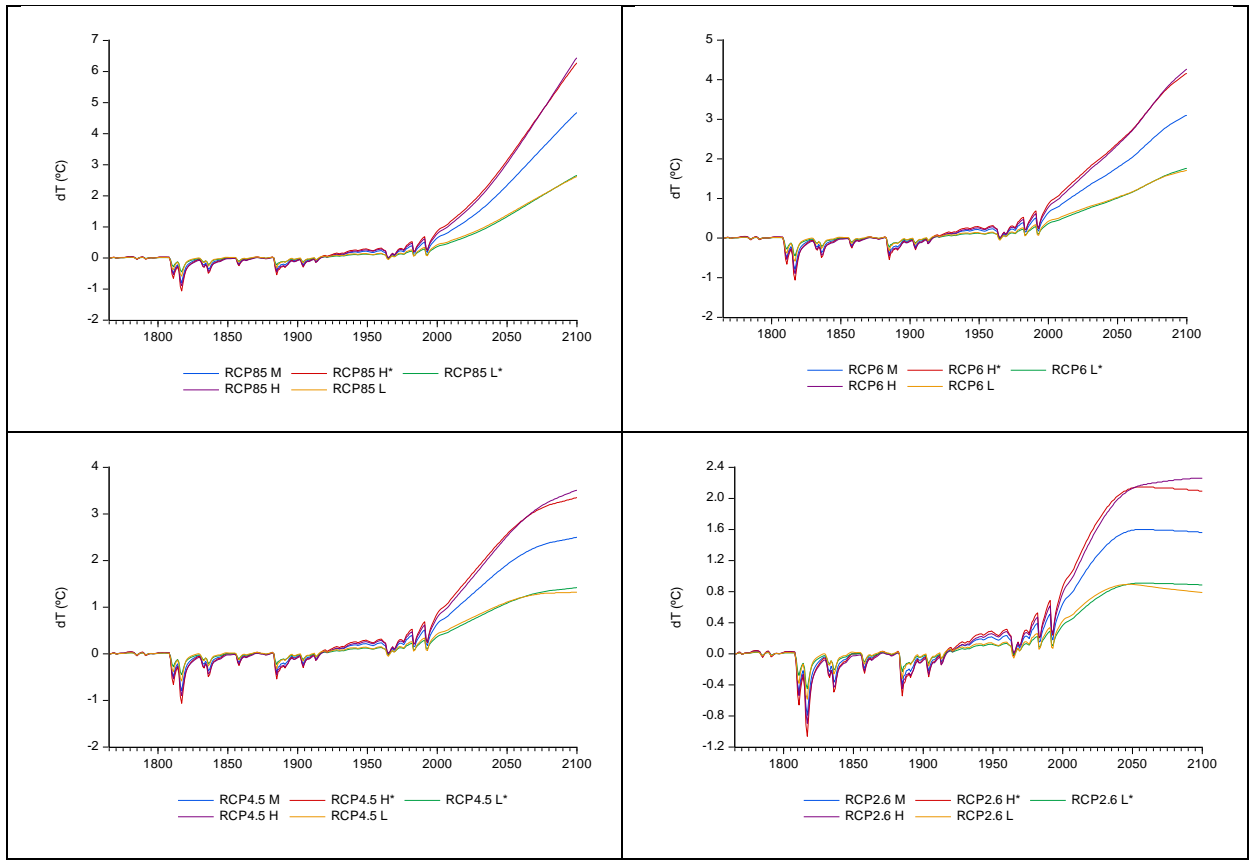

Figure 3. Actual MAGICC6 projections for different RCP scenarios and projections of high/low CS approximated using the parameters in Table 2. M, H, L denote projections using medium, high, and low CS values, while the symbol * indicates the projections are obtained using the average scaling parameters in Table 2.

## 2.3. Regional climate projections

Pattern scaling is a technique for producing regional scenarios based on the robust finding documented in the literature of stationarity in geographical patterns of change in some climate variables, in particular temperature and precipitation (Collins et al., 2013; Santer et al., 1990; Tebaldi and Arblaster, 2014). These patterns can be scaled by global temperature change to produce regional climate change scenarios in a simple and computationally efficient manner that provide a useful approximation to those produced by complex global climate models.

The performance of pattern scaling techniques has been evaluated in the literature in several occasions (Cabré et al., 2010; Collins et al., 2013; Herger et al., 2015; Kravitz et al., 2017a; Mitchell, 2003; Osborn et al., 2018; Tebaldi and Arblaster, 2014; Tebaldi and Knutti, 2018; Zelazowski et al., 2018). These techniques produce adequate approximations for variables such as annual/seasonal temperature and precipitation and other variables excluding extreme events and time-scales in which natural variability is dominant. These patterns are also adequate for most of the emissions scenarios, including low-warming policy trajectories and most regions except where local forcing is strong and time-varying (Collins et al., 2013; Estrada and Botzen, 2021).

The pattern scaling technique can be described as follows (Collins et al., 2013; Estrada and Botzen, 2021; Tebaldi and Arblaster, 2014):

$$P(i,j,t,E,y,s) = T(t,E)p(i,j,y,s) + \xi(i,j,t,s) \qquad (23)$$

where $i$, $j$ denote longitude and latitude, respectively, and $t$ is time. $E$ represents the emissions scenario, $y$ is the climate variable of interest, $s$ is the time of the year for which the scenarios is constructed (annual, month, season) and $T(t)$ is the global annual mean temperature change at time $t$ under the emission scenario $E$. $P(\cdot)$ is the projected field of change for variable $y$ obtained using a complex climate model, while $p(\cdot)$ is the time/emissions scenario invariant spatial pattern of change per 1ºC change in annual global mean temperature, for the climate variable $y$. $\xi(i,j,t,s)$ represents is an error term due to both natural variability and the limitations of the pattern scaling methodology (Collins et al., 2013; Estrada and Botzen, 2021).

The patterns $p(i,j,y,s)$ were calculated for monthly and annual air surface temperature and precipitation using simulations from a battery of coupled ocean-atmosphere general circulation models (Table S2) under the RCP8.5 emissions scenario and that are available at the CMIP5 data portal (https://esgf-node.llnl.gov/search/cmip5/). All simulations were bilinearly interpolated into a common grid with a spatial resolution of 2.5º x 2.5º. The Hodrick Prescott filter (Hodrick and Prescott, 1997) was applied to the time series from each grid point from the climate models' simulations to remove the effects of high frequency variability. Then, temperature/precipitation time series from each grid cell were regressed on the global mean temperature and the slope coefficients were stored as maps that represent the scaling patterns (Kravitz et al., 2017b; Lynch et al., 2017). For producing the regional climate change scenarios, global mean temperature projections from section 2.2 are used to scale the patterns produced in this section. The projections are expressed in ºC for changes in temperatures and in percentage of change for precipitation, with respect to preindustrial conditions (c. 1750). AIRCC-Clim allows the user to select the scaling patters for any given climate model, as well as to generate probabilistic scenarios combining all of them. This stochastic version of AIRCC-Clim uses a uniform distribution which assigns the same probability to each of the climate models. A variety of approaches for assigning probabilities to climate models' output have been proposed in the literature (Knutti et al., 2010; Mendlik and Gobiet, 2016; Xu et al., 2010) but there is no agreement on which would be the best way of doing it (Deser et al., 2014; Knutti, 2010; Notz, 2015; Stephenson et al., 2012). The uniform distribution follows the Principle of Insufficient Reason which is the maximum entropy distribution in absence of any additional information (Gay and Estrada, 2010b; Jaynes and Bretthorst, 2003; Jaynes, 1957). Other probability distributions based on performance evaluation or model dependence could be implemented, however these distributions may be as arbitrary as assigning equal probabilities to each model and may lead to unjustified dismissal of uncertainty (Altamirano del Carmen et al., 2021; Gay and Estrada, 2010b; Potter and Colman, 2003).

Climate change scenarios for temperature and precipitation can be exported as geotiff and netCDF files for three time horizons: 2030 (2021-2040), 2050 (2041-2060) and 2070 (2061-2080). Maps of changes in temperature and precipitation for any year between 2005-2100 can be exported as high-quality PNG and as netCDF files.

### 2.4. Climate risk index generator

AIRCC-Clim uses the probabilistic nature of its projections to produce user-defined risk measures based on thresholds. The current version of the model includes two types of risk metrics: probabilities of exceedance and the dates when the selected threshold is exceeded. The threshold values are selected by the user to reflect his perceptions of risk and information needs.

The calculation of the risks indices is done following the same procedure as CLIMRISK (Estrada and Botzen, 2021). First, the indicator function is used to identify in which simulations and grid cells the threshold is exceeded. In the case of changes in temperature $T$, we have:

$$IRT_{i,j,t,sim} = I(T_{i,j,t,sim} > T^*) \qquad 24$$

where $IRT_{i,j,t,sim}$ is a four-dimensional matrix in which $i,j$ are the longitude and the latitude that define the location of the gird cell, $t$ is time, $sim$ is the number that identifies each of the realizations of the simulation experiment, and $T^*$ is the user-defined threshold in °C. For cases in which the threshold $T^*$ is exceeded the indicator function returns a value of 1 and zero otherwise. In the case of the change in precipitation $P$, the threshold of interest $P^*$ can be positive or negative and thus the indicator function is applied as follows:

$$IRP_{i,j,t,sim} = \begin{cases} I(P_{i,j,t,sim} > P^*) \; if \; P^* > 0 \\ I(P_{i,j,t,sim} < P^*) \; otherwise \end{cases} \qquad (25)$$

$IRP_{i,j,t,sim}$ is a four-dimensional matrix defines as above in which the indicator function returns a value of 1 if the threshold $P^*$ is exceeded and zero otherwise.

Estimates of the probabilities of exceeding the thresholds $T^*$ and $P^*$ can be obtained by summing over the $sim$ dimension:

$$PIRT_{i,j,t} = P(T_{i,j,t} > T^*) = \frac{\sum_{sim=1}^{n} IRT_{i,j,t,sim}}{n} \qquad (26)$$

$$PIRP_{i,j,t} = P(P_{i,j,t} > P^*) = \frac{\sum_{sim=1}^{n} IRP_{i,j,t,sim}}{n} \; if \; P^* > 0 \qquad (27a)$$

$$PIRP_{i,j,t} = P(P_{i,j,t} < P^*) = \frac{\sum_{sim=1}^{n} IRP_{i,j,t,sim}}{n} \; if \; P^* \leq 0 \qquad (27b)$$

Where $PIRT_{i,j,t}$ and $PIRP_{i,j,t}$, are three-dimensional matrices containing probability estimates. The estimated probability maps can be exported as PNG and netCDF files for any year in the period 2005-2100.

These probability estimates of exceeding critical thresholds are used to estimate the expected dates when such thresholds would be attained. These dates can be computed as follows:

$$IDT_{i,j,t} = I(PIRT_{i,j,t} \geq \gamma) \qquad (28)$$

$$IDP_{i,j,t} = I(PIRP_{i,j,t} \geq \gamma) \qquad (29)$$

Where the parameter $\gamma$ represents the percentage of simulations that is required by the user to declare the threshold has been exceeded. $IDT_{i,j,t}$ and $IDP_{i,j,t}$ are matrices in which the entries take the value 1 if the confidence level $\gamma$ is attained or exceeded and zero otherwise. In AIRCC-Clim and CLIMRISK, this value is called *confidence level* and can vary with the risk tolerance of different users. The default value in AIRCC-Clim is 50%. The estimated dates for exceeding the risk threshold are obtained by mapping into a year index the first occurrence of the value 1 in the time dimension of all gird cells in $IDT_{i,j,t}$ and $IDP_{i,j,t}$. The maps of the dates of exceedance can be exported as PNG and netCDF files.

3. Estimating the risks of delaying the implementation of deep mitigation efforts

In this section we provide an example application of AIRCC-Clim in which the effects on climate from a high-emissions trajectory (RCP6.0) are compared to those of a deep mitigation scenario (RCP2.6) consistent with the goals of the Paris Agreement and to those in which such mitigation effort is delayed ten years.

Figure S1 shows the trajectories of $CO_2$, $CH_4$ and $N_2O$ for the RCP6.0, RCP2.6 and the modified version of the RCP2.6 scenarios in which mitigation is delayed for 10 years, starting in 2020. The modified version of the RCP2.6 was edited in Excel and directly loaded to AIRCC-Clim using the user-defined option for emissions scenarios. The RCP6.0 and the RCP2.6 scenarios produce contrasting results in terms of their effects on climate. While the first produces a mean increase in global temperature of about 3°C at the end of this century, and up to 4°C when the 95$^{th}$ percentile is considered, the RCP2.6 limits warming below 2°C for the ensemble mean (about 1.7°C), although this limit is exceeded for the 95$^{th}$ percentile (Figure 4). The MAGICC model produces similar results, with a mean increase in global temperature of 3.1°C and 1.6°C for the end of the century under the RCP6.0 and RCP2.6 scenarios, respectively (Figure S2). The TCM produces a considerably larger mean increase for the RCP6.0 (3.9°C), but that still lies within the likely range of the CMIP5 experiment (Table 1). The ST simulations for the modified RCP2.6 in which the mitigation effort is delayed for 10 years, show that the mean increase in global temperature reaches 2°C in 2100, and up to 3°C in the 95$^{th}$ percentile.

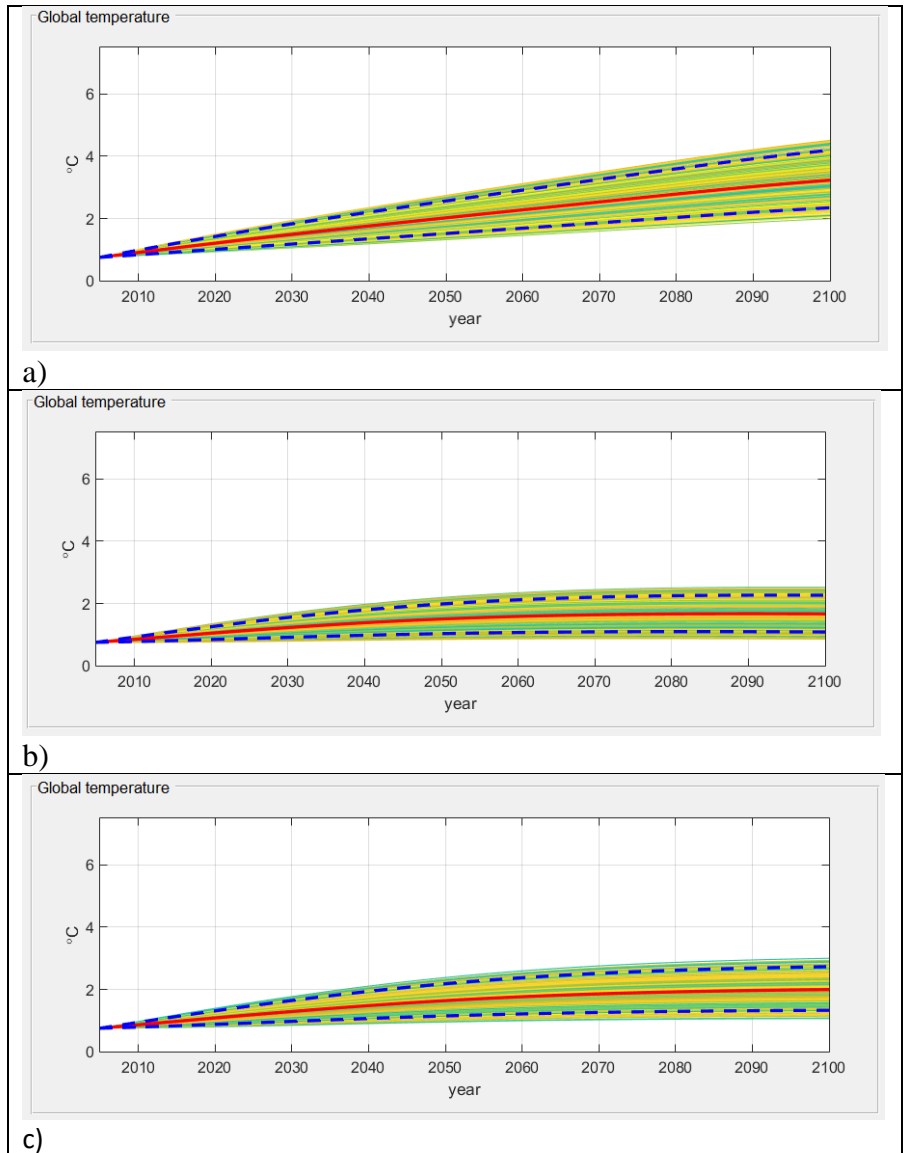

Figure 4. Global temperature projections using the ST model. Panels a), b) and c) show the simulations for the RCP6.0, RCP2.6 and the modified RCP2.6 scenarios, respectively. The red bold line shows the mean of the ensemble while the dark slashed blue lines depict the 5$^{th}$ and 95$^{th}$ percentiles of the ensemble, and the light green and yellow lines show the individual simulations. Each experiment has 500 simulations.

Figures 5 and 6 show the changes in annual mean temperature and total annual precipitation at the grid cell level for the three emissions scenarios and two time-horizons (2050 and 2100). Temperature increase is much larger in high latitudes of the northern hemisphere due to the Arctic Amplification phenomenon (Pithan and Mauritsen, 2014), reaching about 6ºC for the mid-century and more than 8ºC in 2100, under the RCP6.0 scenario (see Figure S3 and S4 for results using the precalculated MAGICC6 and TCM runs). Most of the continents would experience temperature increases in 2050 of about 2ºC to 3ºC and of about 4ºC to 5ºC in 2100. Apart from the Arctic region, midlatitudes in North America and in Eurasia would have the largest increases (5ºC-6ºC) in temperature by the end of the present century, followed by Southern Asia and the Middle East, North and South Africa, parts of the west coast of North America, Mexico and the Amazonian region and the northern part of Brazil. Large changes are also expected in precipitation under the RCP6.0 scenario, with large increases in high latitudes, the equatorial Pacific Ocean and some parts of the Middle East, and large decreases in the Mediterranean, the Caribbean, Mexico, and the southern part of the US, as well as in southern parts of Africa and America (Figure S3 and S4). The implementation of a deep mitigation effort consistent with the goals of the Paris Agreement would significantly limit these changes in climate. Under the RCP2.6 most of the Arctic would not exceed a warming of 5ºC during this century, most continents would not exceed 3.5ºC, and precipitation change would be notably smaller.

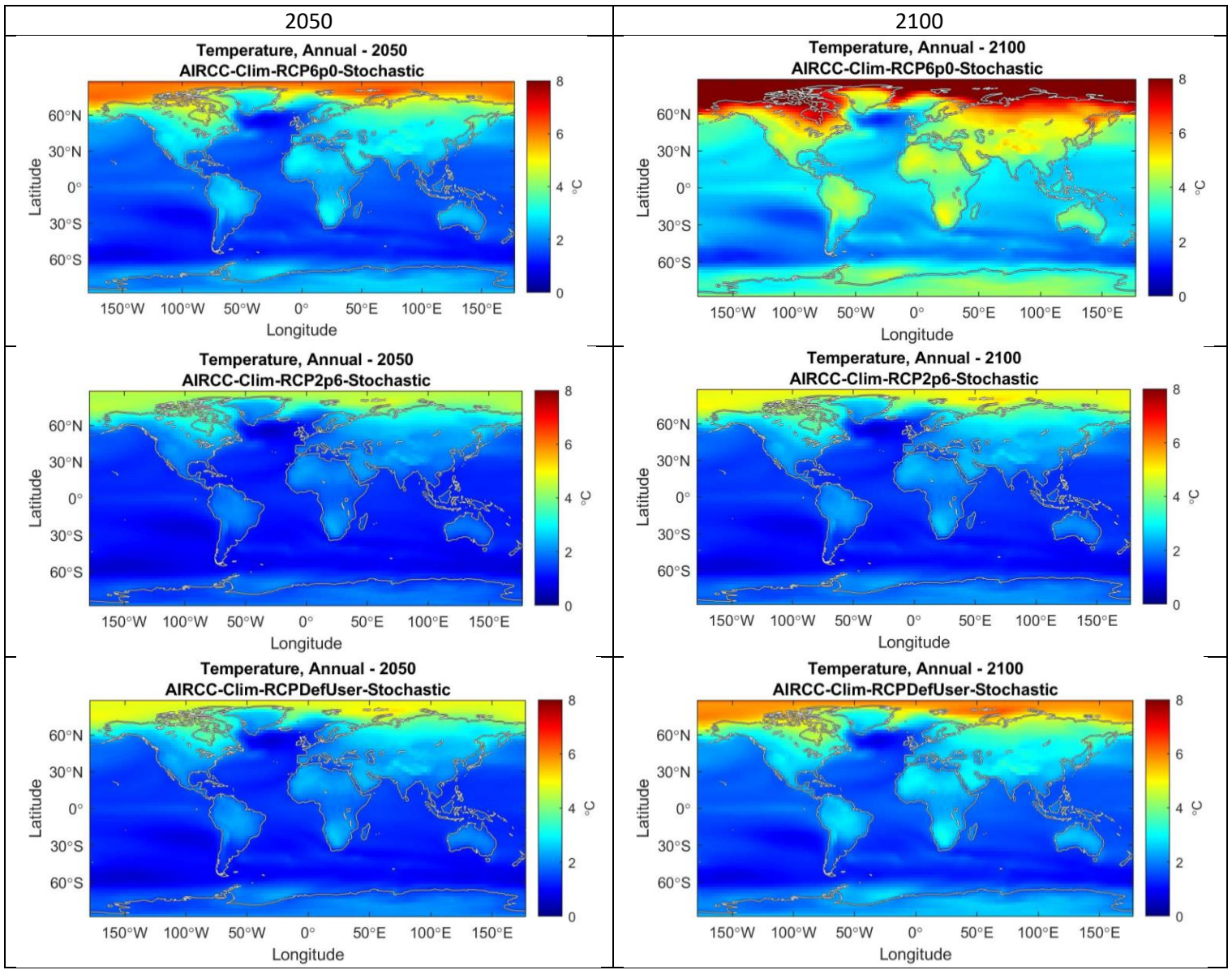

Figure 5. Annual temperature change projections (ºC) for different emissions scenarios estimated by the modified ST model. The upper panel shows the changes in temperature under the RCP6.0 scenario for 2050 (left) and 2100 (right). The middle panel shows the changes in temperature under the RCP2.6 scenario for 2050 (left) and 2100 (right). The lower panel shows the changes in temperature under the delayed RCP2.6 scenario for 2050 (left) and 2100 (right).

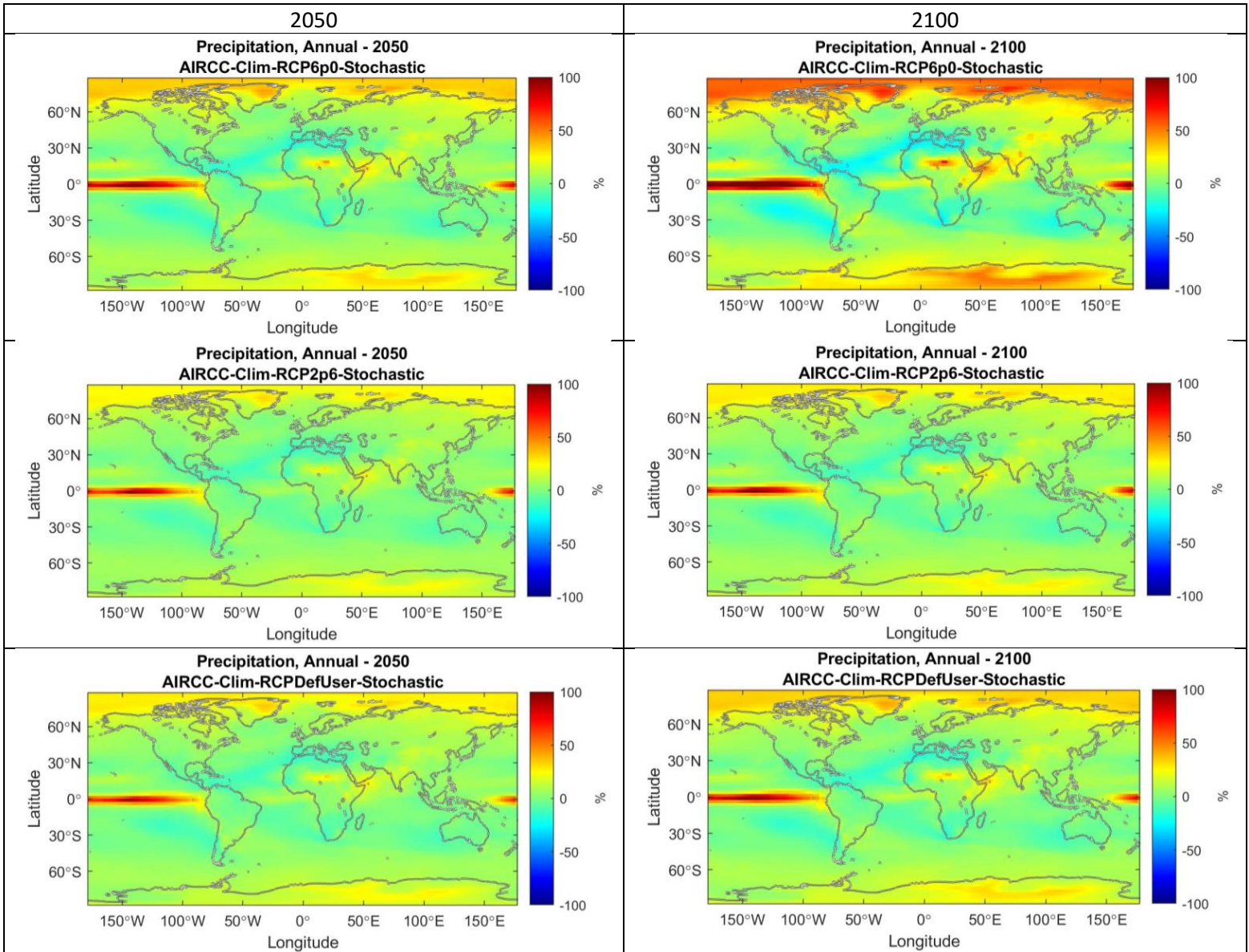

Figure 6. Annual precipitation projections (%) for different emissions scenarios estimated by the modified ST model. The upper panel shows the changes in precipitation under the RCP6.0 scenario for 2050 (left) and 2100 (right). The middle panel shows the changes in precipitation under the RCP2.6 scenario for 2050 (left) and 2100 (right). The lower panel shows the changes in precipitation under the delayed RCP2.6 scenario for 2050 (left) and 2100 (right).

AIRCC-Clim portraits the risks of climate change, and the benefits of mitigation, in a clearer way due to its probabilistic nature and its capacity to produce maps of probabilities and of dates of exceedance. Figures 7 and 8 show the probabilities of exceeding a 2.5ºC increase in annual temperatures and a decrease of 15% in annual precipitation in 2050 and 2100. Under the RCP6.0, by 2050 the probabilities of exceeding 2.5ºC increase in annual temperatures are higher than 60% for most of the continents, while the Arctic region and the high latitudes of the northern hemisphere are virtually certain to exceed this threshold by mid-century (Figure S5). These simulations also show that a 2.5ºC warming would be exceeded in all continents by 2100 and that there is a high probability (above 60%) that this threshold would be crossed in all oceans, except for parts of the southern hemisphere and the Atlantic. The probabilities of exceeding a decrease of 15% in total annual precipitation in 2050 are higher than 50% for several areas of the world such as the Mediterranean, parts of Northern, Western and Southern Africa, the Caribbean and Mexico (Figure S5). For 2100, these probabilities increase to more than 60% in those regions and extend to parts of South America and West Australia. If the emissions trajectory described in by the RCP2.6 is achieved, by 2050 the probability of exceeding 2.5ºC would be lower than 40% over the continents, except for the Arctic and parts of the midlatitudes in the northern hemisphere where the probabilities range from 60% to 100% (Figure 7). Likewise, the probabilities of exceeding decreases in precipitation of at least 15% are considerably smaller in comparison to the RCP6.0. However, by the end of this century the probabilities of exceeding this threshold would be close to 50% for the southern region of Spain, parts of North and West Africa (Morocco, Mauritania, Mali, Senegal, Sierra Leone and Guinea), and about 40% for parts of the Caribbean, Central and South America (Colombia and Venezuela; Figure 8).

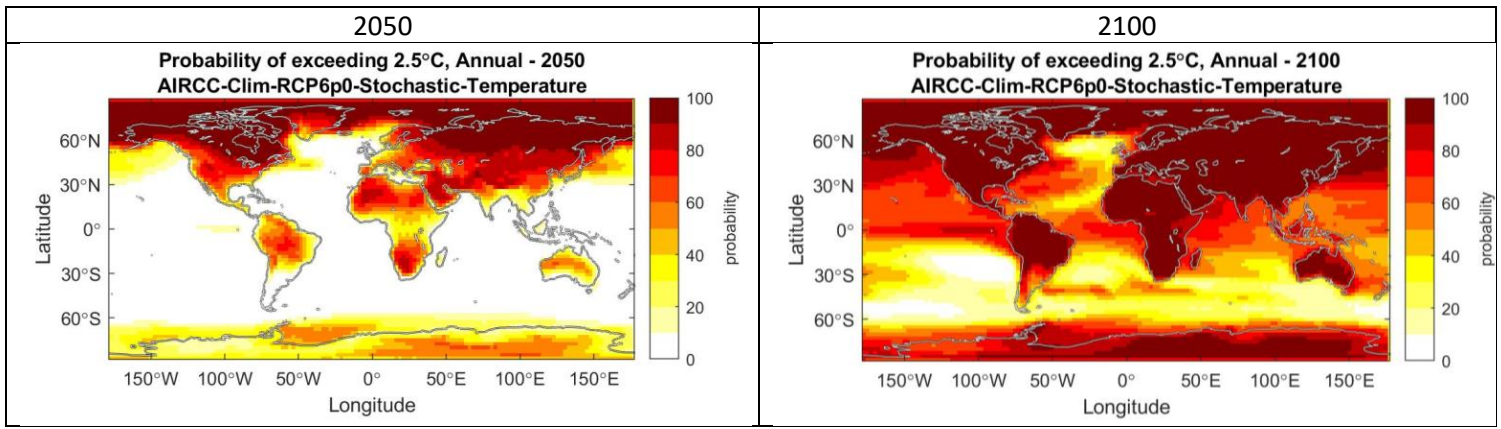

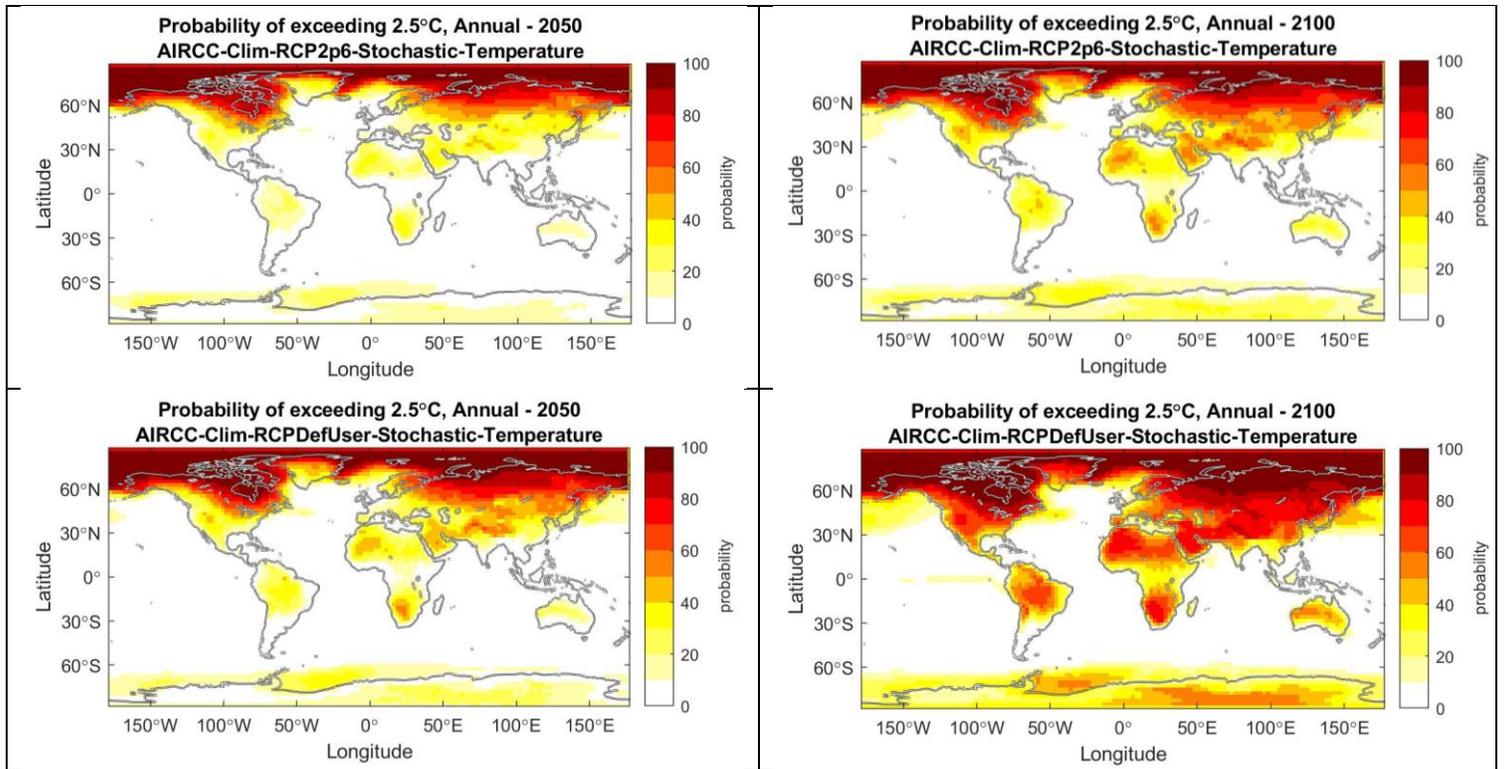

Figure 7. Probabilities of exceeding increases of 2.5°C in annual temperature for different emissions scenarios estimated by the modified ST model. The upper panel shows the probabilities of exceedance for the RCP6.0 scenario in 2050 (left) and 2100 (right). The middle panel shows the probabilities of exceedance for the RCP2.6 scenario for 2050 (left) and 2100 (right). The lower panel shows the probabilities of exceedance for the delayed RCP2.6 scenario for 2050 (left) and 2100 (right). Probabilities are expressed in percentages.

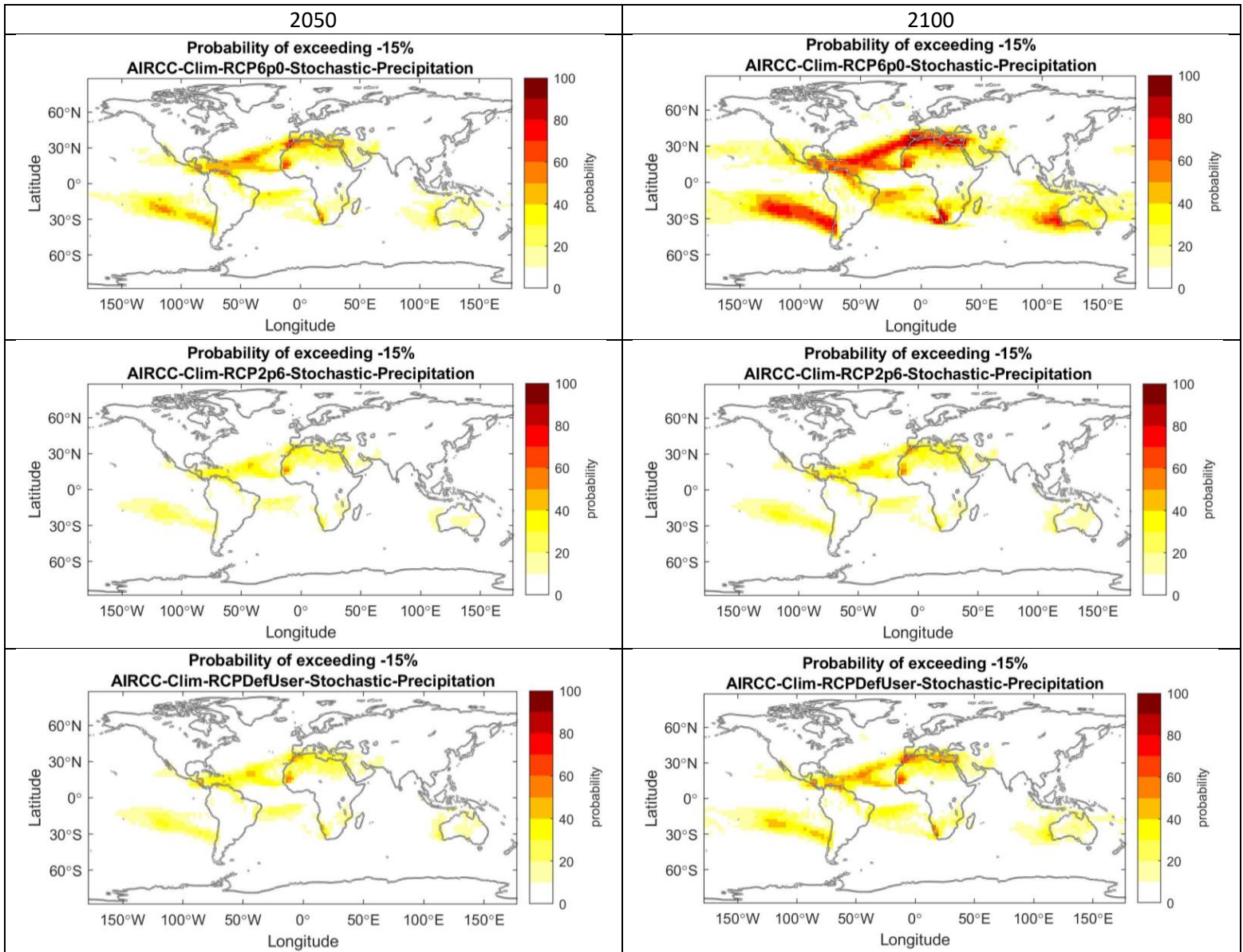

Figure 8. Probabilities of exceeding decreases of 15% in annual precipitation for different emissions scenarios estimated by the modified ST model. The upper panel shows the probabilities of exceedance for the RCP6.0 scenario in 2050 (left) and 2100 (right). The middle panel shows the probabilities of exceedance for the RCP2.6 scenario for 2050 (left) and 2100 (right). The lower panel shows the probabilities of exceedance for the delayed RCP2.6 scenario for 2050 (left) and 2100 (right). Probabilities are expressed in percentages.

Figure 9 shows the dates when the 2.5ºC threshold would be exceeded. The default confidence level ($\gamma = 50$) was used for the estimates presented in this section. The results show that under the RCP6.0, most of the planet except for part of the southern oceans and part of the North Atlantic, would exceed the 2.5ºC threshold during this century and that some parts of the world already exceeded it (Figure S5). The date of exceedance for the Arctic occurred during the 2000s, while for much of the high latitudes in the northern hemisphere, the Middle East, parts of South Asia, and West and South Africa, the exceedance is expected to occur in the 2020s-2030s. The remainder of the continents and the Antarctic region would go over this threshold during the period 2040-2060 and most of the oceans above the 20ºS would exceed the 2.5ºC threshold during this century. The regions in which reductions of at least 15% in annual precipitation is exceeded are fewer and form well-defined geographical patterns that cover the Mediterranean, parts of North, West and South Africa, Central America and the Caribbean, Mexico and Colombia and Venezuela in South America, as well as the west part of Australia and part of the southern Pacific Ocean (Figures 9 and S5). The dates for exceedance on these regions are typically reached in the 2050-2060 decades, although regions of Spain and West Africa could exceed this threshold as early as the 2040s.

Achieving RCP2.6 would prevent exceeding these thresholds for most of the world during this century (Figure 9). However, as mentioned above, some regions such as the Arctic and the high latitudes of the northern hemisphere already have exceeded the 2.5ºC temperature threshold or will do so during the next decade, regardless of the emission scenario that is selected. For parts of the midlatitudes, the RCP2.6 represents delaying reaching the 2.5ºC threshold for about 20 years, which buys time for adapting to the projected changes and to reduce risks and damages. The occurrence of this threshold would also be delayed until 2060 in some parts of the Sahara, South Africa, the Middle East and India. Exceeding decreases in precipitation of more than 15% would not occur during this century, with the exception of a few grid cells in Africa, Spain and Central America.

AIRCC-Clim runs illustrate that delaying the deep mitigation efforts of the RCP2.6 by ten years would significantly increase the risks of climate change during this century. The bottom rows of Figures 7 and 8 show that, under delayed international action, the probabilities of exceeding 2.5ºC in annual temperature and a decrease of at least -15% annual precipitation in 2100 are similar to those obtained in 2050 for the RCP6.0 and much higher than those of the original RCP2.6. The consequences of delaying for ten years the mitigation efforts described in the RCP2.6 are clearly illustrated by Figure 9: North and South Africa, South America, the Middle East and Central Asia would exceed a 2.5ºC increase in temperatures around 2060, when this threshold was not reached during this century under the original RCP2.6 trajectory. Parts of Australia would exceed this threshold at the end of the present century if mitigation efforts were postponed. In terms of precipitation reductions, the Mediterranean region would be most affected, as the dates of exceedance of decreases of at least 15% in precipitation would occur as soon as 2040-2050 for the south of Spain and North Africa and at the end of this century for Greece. Western Africa would go over this threshold in 2040, and parts of Central America and South Africa would see decreases of more than 15% in the 2050-2060 period. However, it is important to note that this delayed action scenario still provides important benefits in comparison with the RCP6.0, which is commonly used to represent current international mitigation commitments. Some of the most affected regions due to warming would buy time (about 2 decades) for adapting to a 2.5ºC under the delayed version of the RCP2.6. This is not so clear with regard to exceeding -15% decrease in annual precipitation for the most affected regions, as in comparison with the RCP6.0, the delayed version of the RCP2.6 would buy them only 5-10 years for implementing adaptation actions.

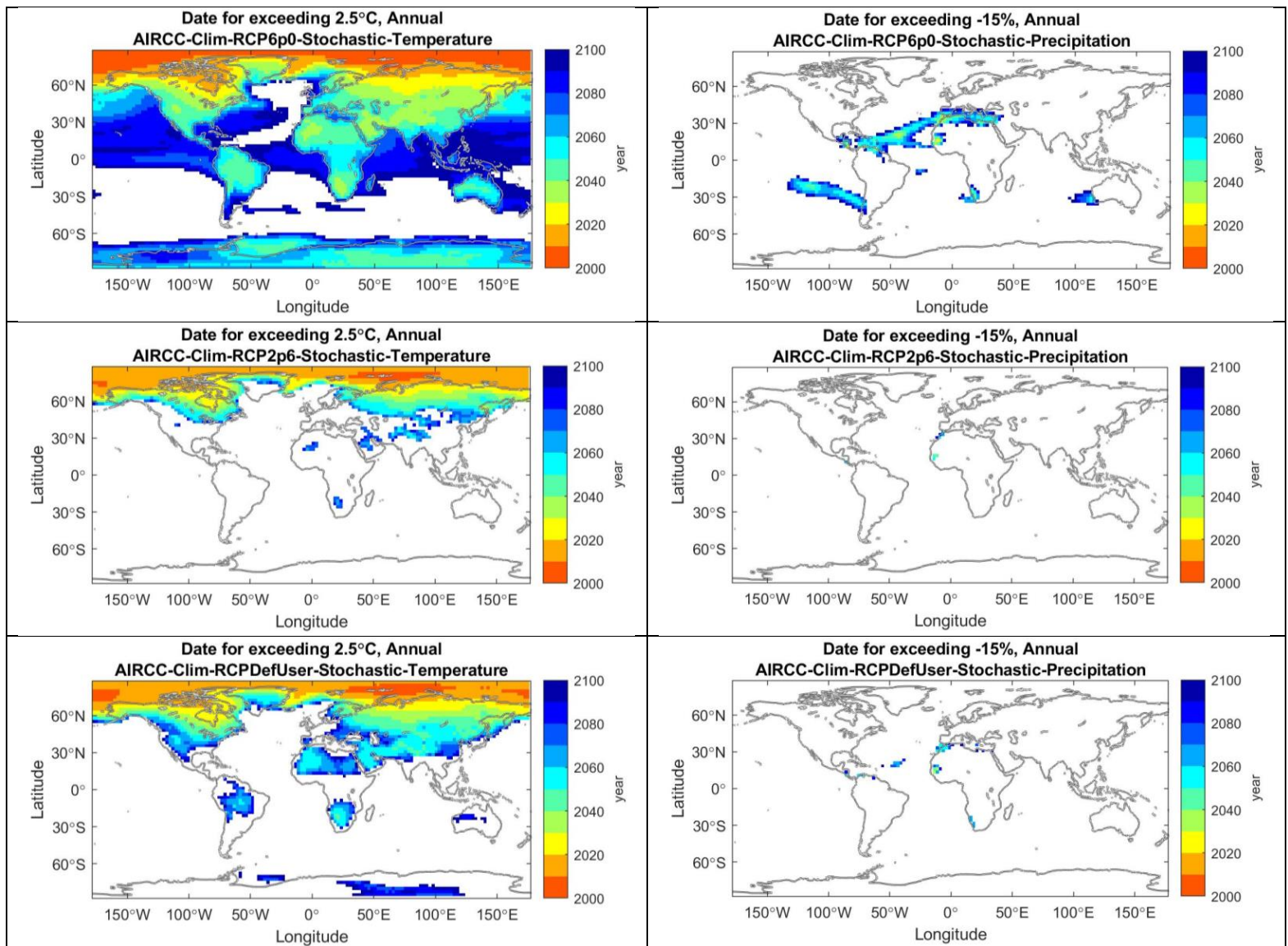

Figure 9. Dates of exceedance for increases of 2.5°C in annual temperature and decreases of 15% in annual precipitation for different emissions scenarios estimated by the modified ST model. The upper panel shows the dates of exceedance for annual temperature (left) and annual precipitation (right) under the RCP6.0 scenario. The middle panel shows the probabilities of exceedance for annual temperature (left) and annual precipitation (right) under the RCP2.6 scenario. The lower panel shows the probabilities of exceedance for annual temperature (left) and annual precipitation (right) under the delayed RCP2.6 scenario.

Monthly estimates of changes in precipitation and temperature are commonly needed for assessing the impacts of climate change in natural and human systems. AIRCC-Clim also generates estimates of monthly temperature and precipitation change, as well as estimates of probabilities and dates of exceedance for user-defined thresholds. Figure 10 illustrates this feature for the RCP6 emissions scenario and for the central month of winter and summer (i.e., February and July).

During the coldest months in the northern hemisphere's winter, the threshold of 2.5ºC was exceeded at the beginning of this century in the Arctic, while for parts of the midlatitudes it will be exceeded during this decade of in the 2030s (Figure 10a). In most of the remaining parts of the northern hemisphere the 2.5ºC threshold in temperatures during February would be exceeded in the 2030-2050 decades. For parts of North and Central America, the driest months occur in winter and precipitation in February in those areas would decrease at least 15% as soon as 2030. In the case of regions in southern hemisphere, such as Australia, Central and South Africa, as well as most of South America, the temperature threshold during one of the hottest months (February) would occur before 2060 and, in some parts of these region, this threshold could be exceeded 20 years earlier.

The hottest months in the northern hemisphere occur during the boreal summer. As shown in Figure 10c, exceeding the threshold a 2.5ºC increase in July would happen in this decade for regions in the Mediterranean such as Spain, France, Italy, Greece, and parts of North Africa. In these regions, exceeding this threshold in temperatures during July would be accompanied by decreases of at least 15% in precipitation before 2050 in the same month, which is one of the driest in the Mediterranean.

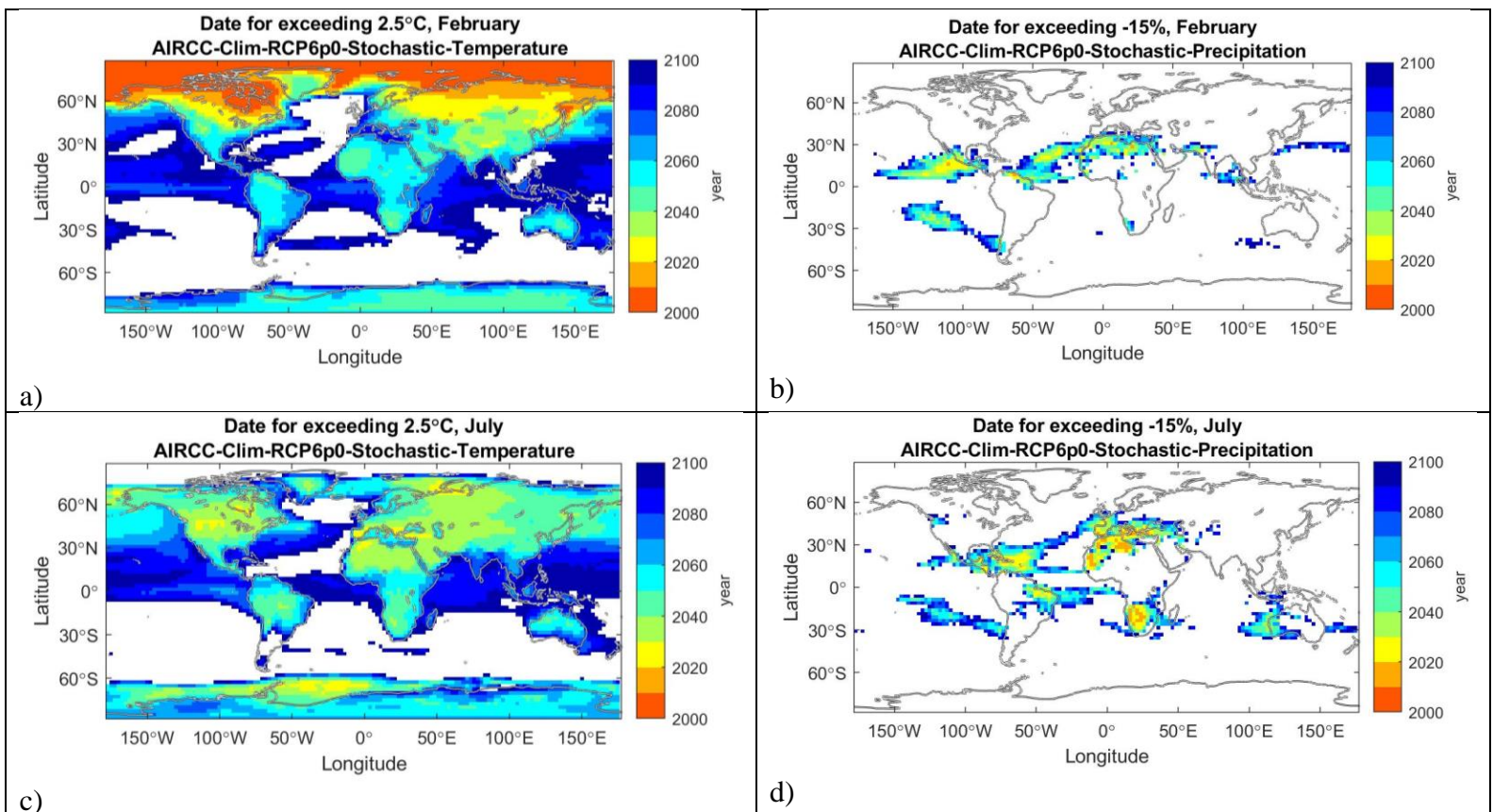

Figure 10. Estimates of dates of exceedance during February and July for 2.5ºC and -15% thresholds in temperature and precipitation, respectively.

Moreover, due to its capabilities for exporting output, AIRCC-Clim can be easily combined with other products to address the user's specific information needs. Figure 11 combines population count projections from the SSP3 scenario that were obtained from CLIMRISK (Estrada and Botzen, 2021) with two risk measures produced with AIRCC-Clim to provide a first approximation of risk and exposure. Climate and population projections show that by 2050 some regions of the world will have high exposure and high probabilities of experiencing large changes in climate. In the bivariate map shown in Figure 11a dark magenta color indicate regions for which large population and high probabilities of exceeding 15% decrease in precipitation are projected. These high-risk, high-exposure regions include large fractions of the Mediterranean, Central America and parts of the Middle East and South Asia. Light yellow areas indicate regions characterized by large population but low probabilities of decreases in precipitation of at least 15%. These include high latitude regions in the northern hemisphere for which most climate models' projections suggest an increase in precipitation. This combination of population and precipitation change would be found in parts of India, China, parts of central, northern and eastern Europe, northern US and Canada. Light blue regions such as Australia, large parts of Noth Africa and South America, are where decreases in precipitation of at least 15% are highly likely but where population counts are low.

Figure 11b shows a bivariate map of population counts and the probability of exceeding 2.5ºC in annual temperature change by year 2050. Regions such as the eastern part of the US, Central America, most of Europe, India, China, the Middle East and parts of Africa are shown in dark magenta color. These regions are characterized by high probabilities of exceeding 2.5ºC and large population counts. Regions in light blue hue represent places where the probability of exceeding a warming of at least 2.5ºC in 2050 are high, but population in those areas is not large. This is the case of high latitudes in the northern hemisphere, Australia, the Amazon rainforest, the Sahara, Namib and the Arabian deserts. Moreover, Figures 11 helps to identify risk hotspots in which population counts will be high in the future and significantly dryer and hotter conditions will likely occur. Such combination of factors has been associated with higher risks of human conflict and migration (Barrios et al., 2006; Hodler and Raschky, 2014; Hsiang et al., 2013; Puente et al., 2016; World Bank, 2016), as well as impacts on biomass production and more frequent wildfires (De Dato et al., 2008; Stevens-Rumann et al., 2018).

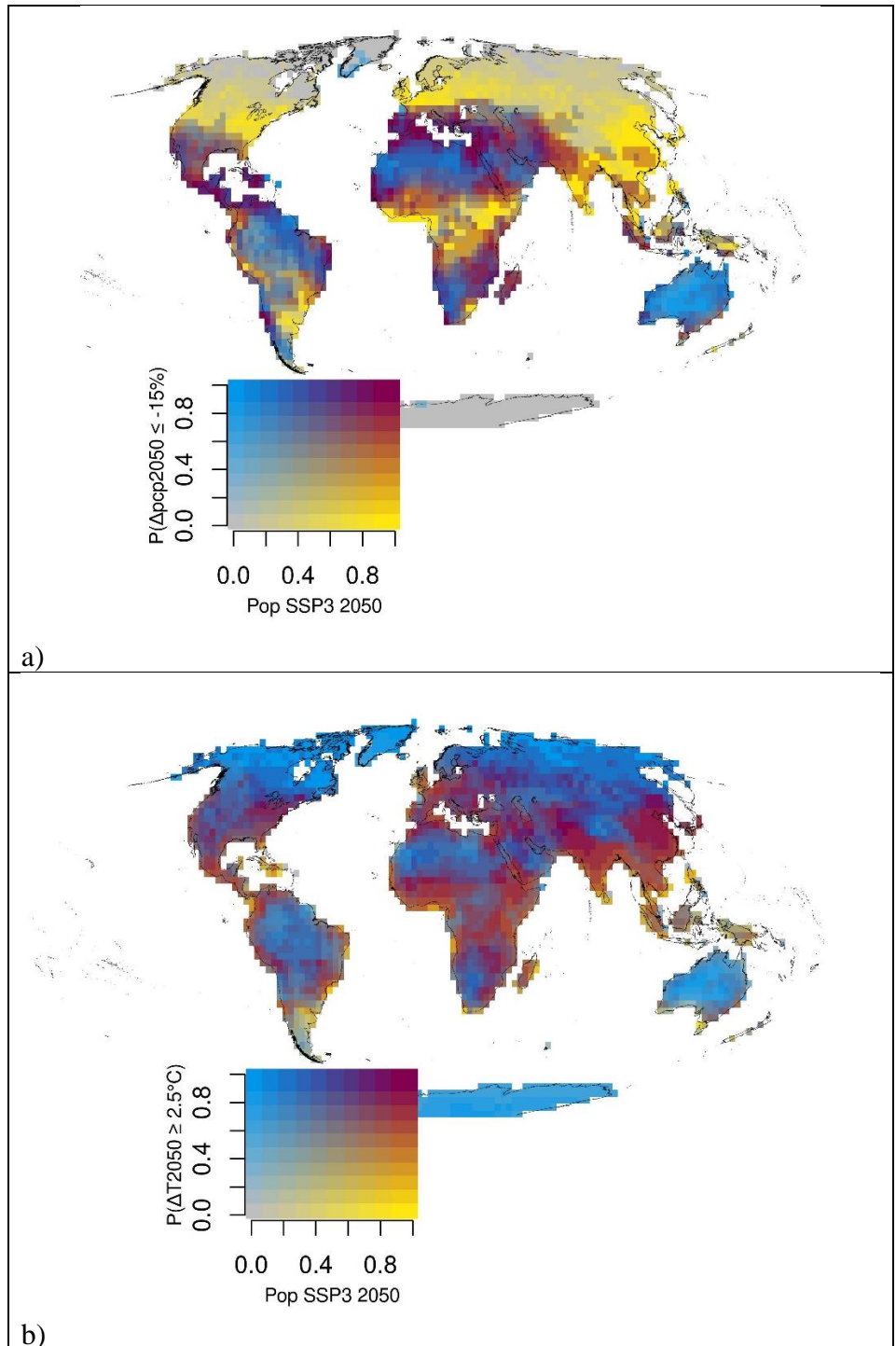

Figure 11. Bivariate maps of population and probabilities of user-defined risk thresholds. Panel a) shows a bivariate map of population counts in year and of the probabilities of decreases in annual precipitation of at least 15% in year 2050. Panel b) shows a bivariate map of population counts and of the probabilities of warming of at least 2.5°C in year 2050.

4. Conclusions

Here we present AIRCC-Clim, an emulator of complex climate models included in the IPCC's Fifth Assessment Report that allows generating probabilistic climate change projections and risk measures for RCP emissions scenarios, as well as for user-defined emissions scenarios. Global temperature projections are produced using a modified version of the ST model and precalculated runs of the MAGICC and TCM models. AIRCC-Clim has a spatial resolution of 2.5º x 2.5º and produces monthly and annual temperature and precipitation scenarios. This is a user-friendly, stand-alone software aimed for students, decision-makers, and researchers that allows for quick estimates of changes in climate, as well as of the probabilities and dates of exceedance of user-defined thresholds. The AIRCC-Clim model attempts to fill users' needs for models that have low technical and computing requirements, but that are able to emulate complex climate models' output and produce spatially explicit, probabilistic projections and risk measures.

AIRCC-Clim extends the ST climate model to include a dynamic climate sensitivity that takes advantage of the well-established approximately linear relationship between cumulative $CO_2$ emissions and global temperature increase. This extension of the ST model, combined with stochastic simulation, allows to closely approximate the best estimate and likely range included in the Fifth Assessment Report of the IPCC. By means of a simple stochastic simulation procedure, we account for the uncertainty in the climate sensitivity parameter and produce probabilistic scenarios based on MAGICC and TCM precalculated runs. Extensions and future development of this model include the integration with IVA and integrated assessment models, such as simple agricultural emulators and climate-economy models (Estrada et al., 2020; Estrada and Botzen, 2021; Ignjacevic et al., 2021, 2020); the inclusion of additional climate variables (e.g., minimum and maximum temperatures and bioclimatic indices), as well as complementary uni- and multivariate risk measures.

Acknowledgements: Francisco Estrada acknowledges financial support from DGAPA-UNAM through the projects PAPIIT IN110718 and IN111221 and from PINCC-UNAM.

*Supplementary Figures*

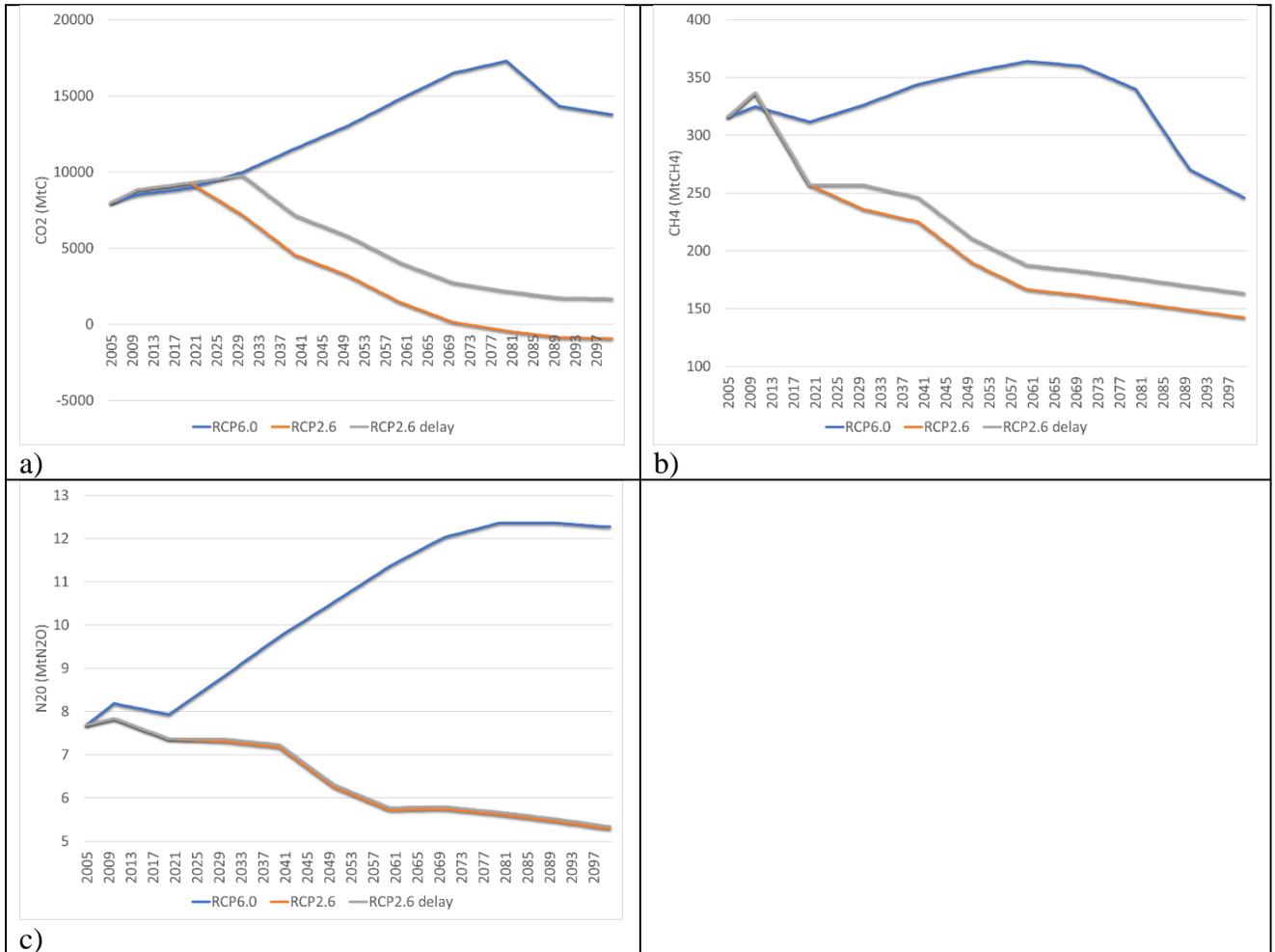

a)

b)

c)

Figure S1. Emissions scenarios for the RCP6.0, RCP2.6 and a modified version of the RCP2.6 in which mitigation efforts are delayed for 10 years, starting in 2020. Panel b) shows the trajectories of emissions of $CO_2$ for the RCP6.0 (blue), RCP2.6 (orange) and the modified RCP2.6 (grey). Panels b) and c) show the corresponding emissions trajectories for $CH_4$ and $N_2O$, respectively.

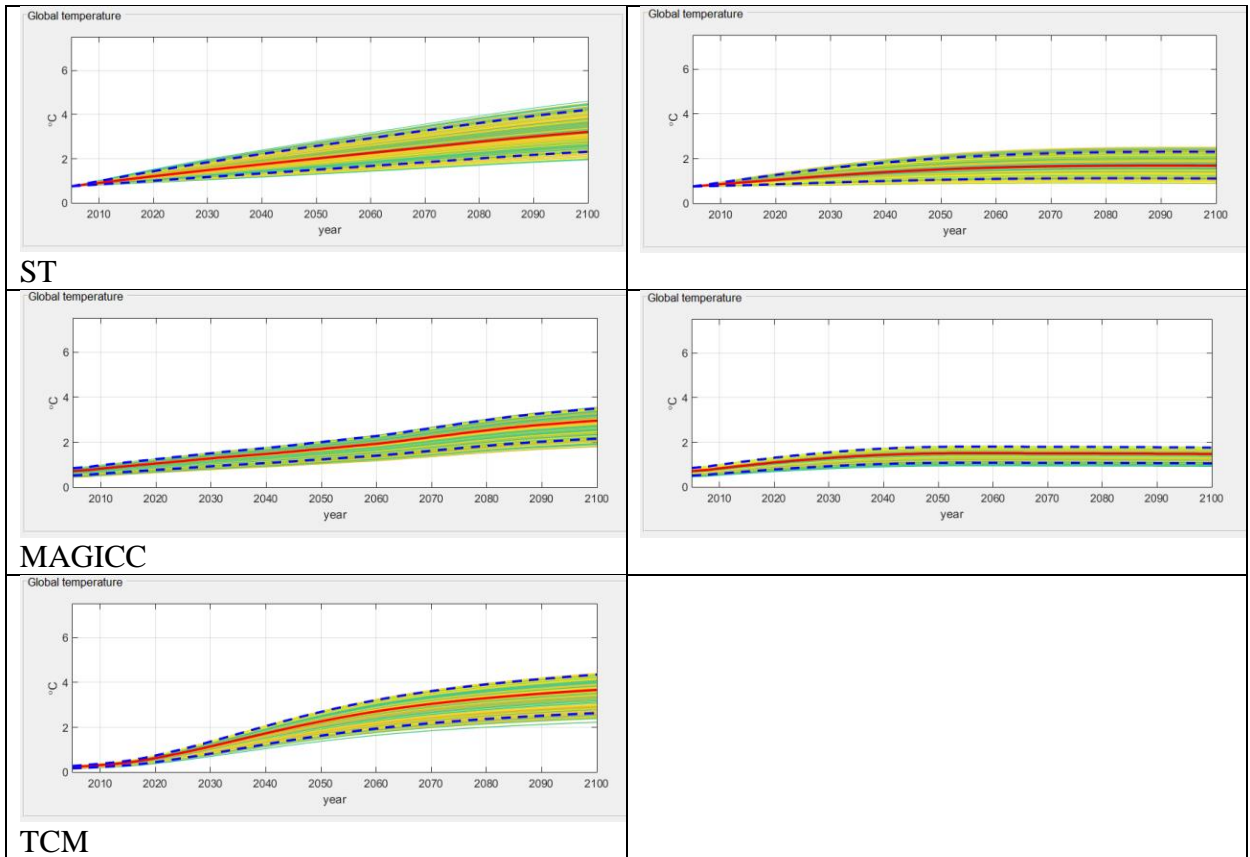

Figure S2. Global temperature projections under the RCP6.0 and RCP2.6. The left and right columns show the projections for the RCP6.0 and the RCP2.6, respectively. The upper row shows the simulations of the ST model, while the middle and bottom rows show the projections based on the MAGICC and TCM model runs, respectively.

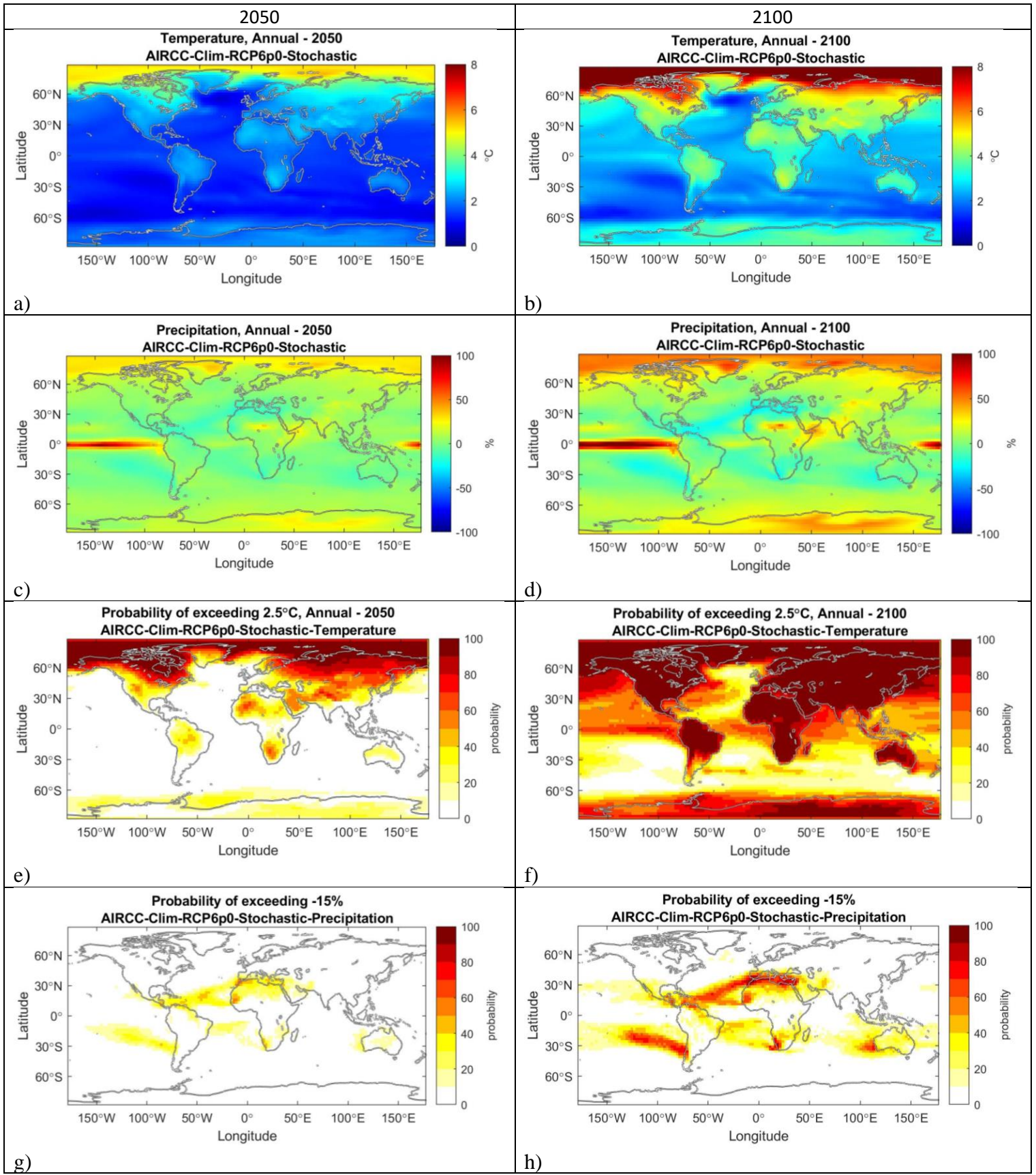

Figure S3. Projections based on MAGICC6 runs for annual temperature and precipitation. Panel a) and b) show the changes in annual temperature under the RCP6.0 for 2050 and 2100, respectively. Panels c) and d) show the changes in annual precipitation under the RCP6.0 for 2050 and 2100, respectively. Panels e) and f) show the probabilities of exceeding 2.5ºC increase in annual temperature under the RCP6.0 for 2050 and 2100, respectively. Panels g) and h) show the probabilities of exceeding decreases of -15% in annual precipitation.

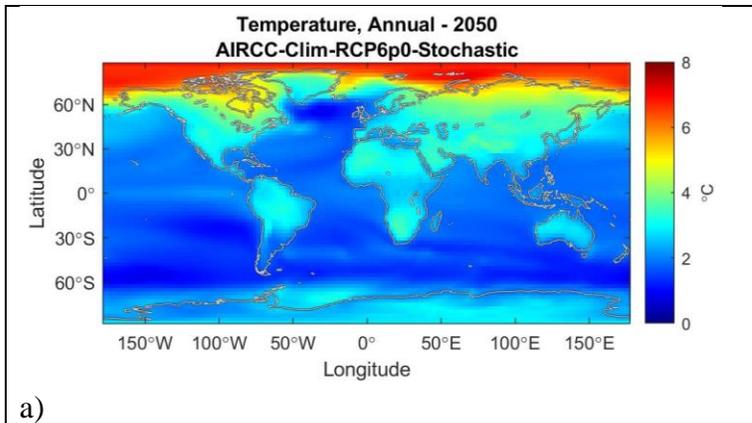 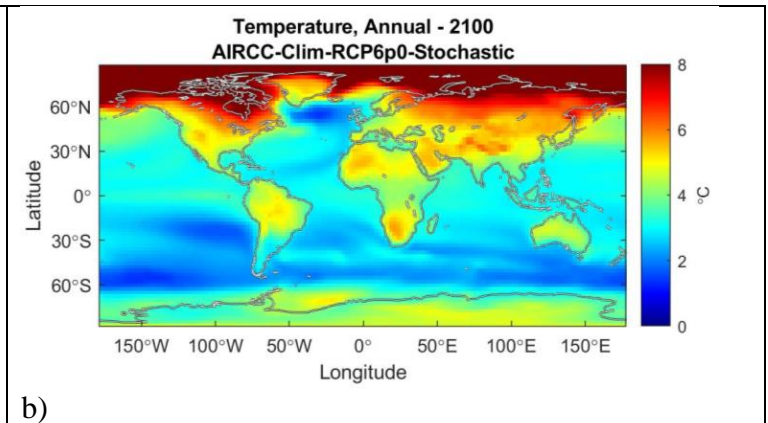
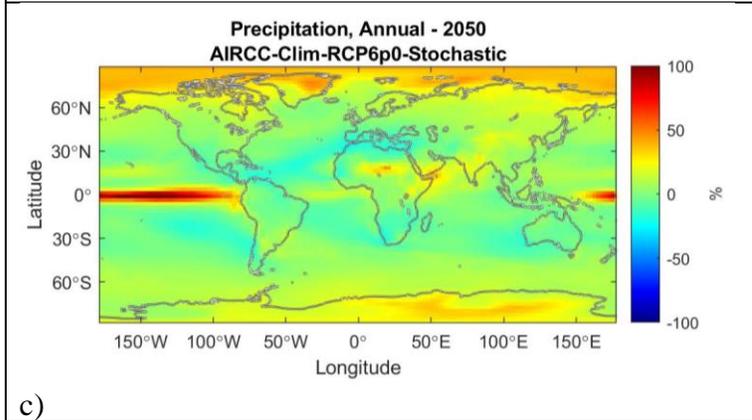 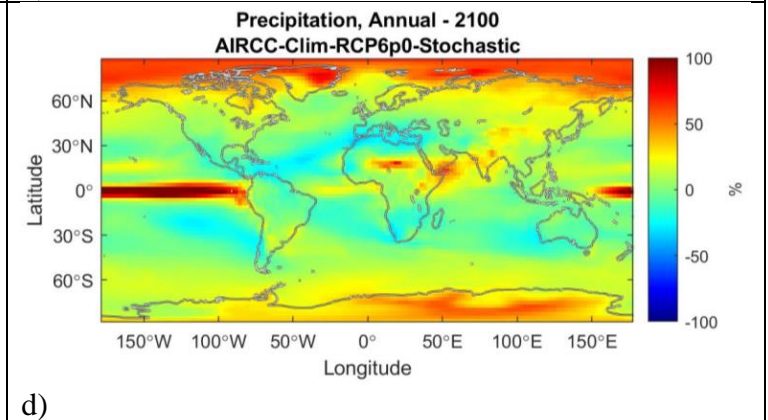
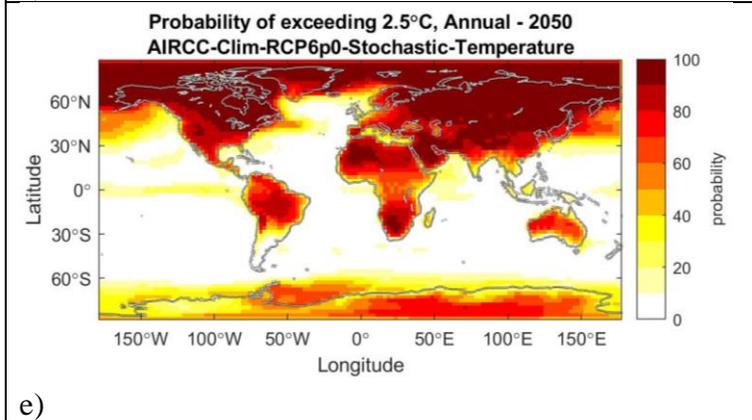 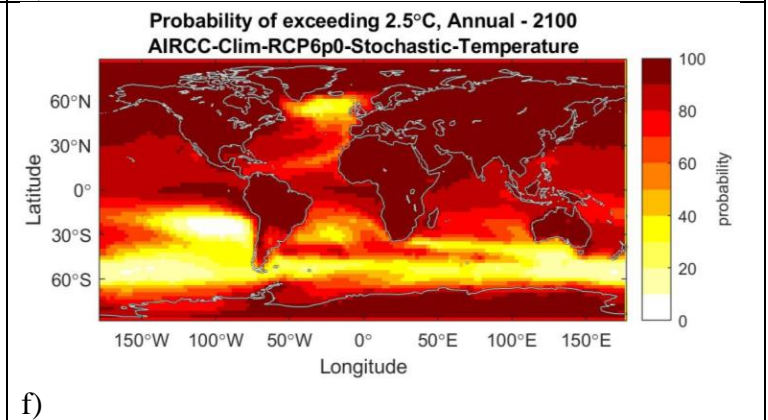
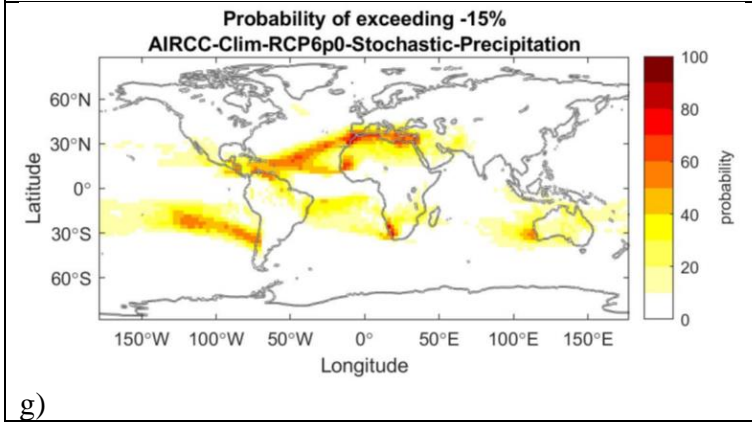 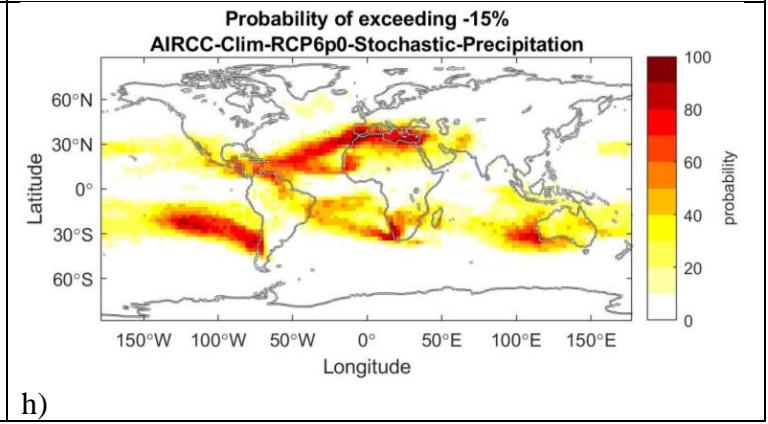

Figure S4. Projections based on TCM runs for annual temperature and precipitation. Panel a) and b) show the changes in annual temperature under the RCP6.0 for 2050 and 2100, respectively. Panels c) and d) show the changes in annual precipitation under the RCP6.0 for 2050 and 2100, respectively. Panels e) and f) show the probabilities of exceeding 2.5ºC increase in annual temperature under the RCP6.0 for 2050 and 2100, respectively. Panels g) and h) show the probabilities of exceeding decreases of -15% in annual precipitation.

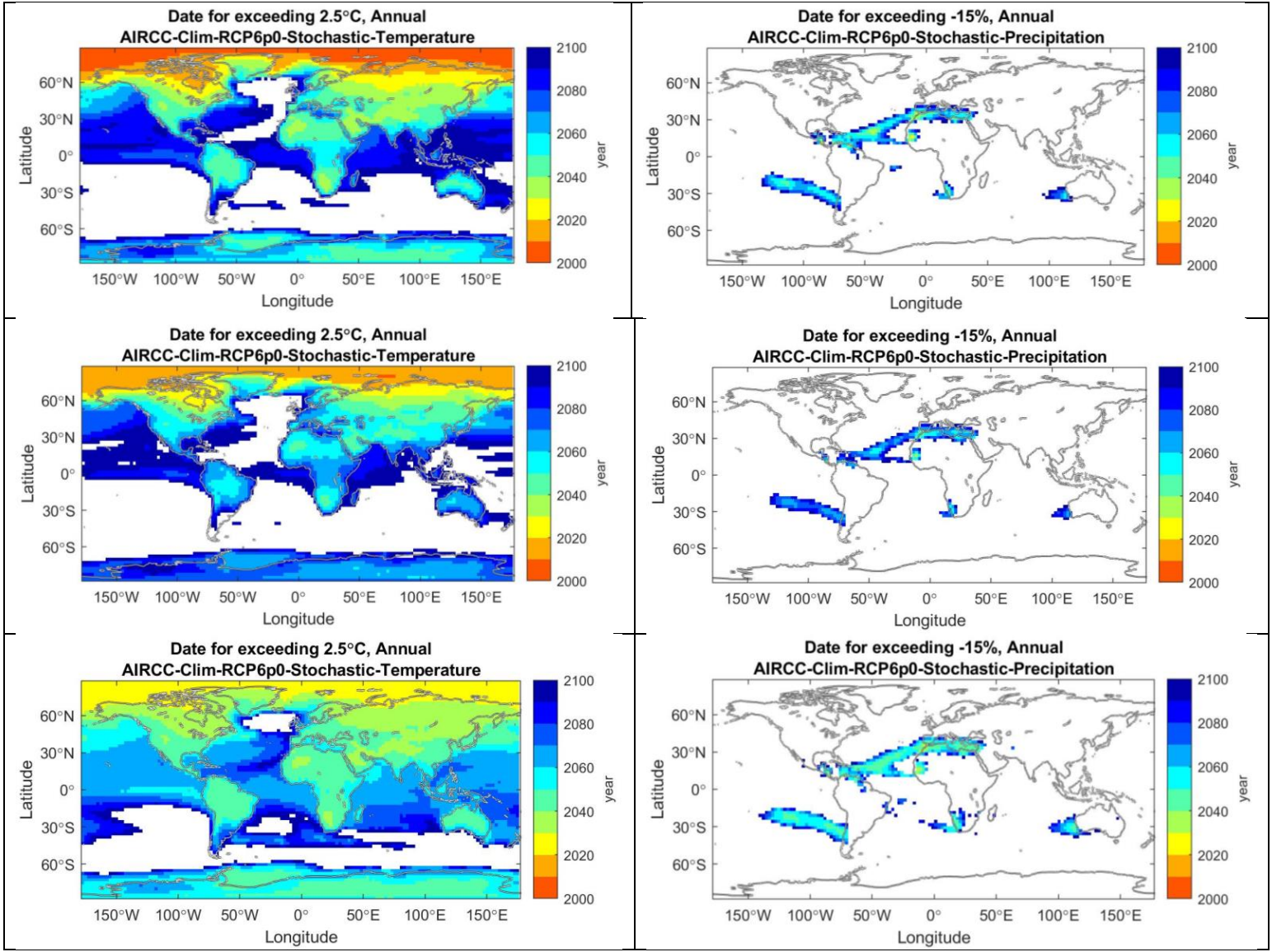

Figure S5. Dates of exceedance for annual temperature and precipitation. The upper panels show the dates for exceeding a 2.5ºC increase in annual temperature (left) and the dates for exceeding a 15% decrease in annual precipitation (right), for the RCP6.0 estimated by the modified ST model. The middle panels show the dates for exceeding a 2.5ºC increase in annual temperature (left) and the dates for exceeding a 15% decrease in annual precipitation (right), for the RCP6.0, based on the precalculated MAGICC6 runs. The lower panels show the dates for exceeding a 2.5ºC increase in annual temperature (left) and the dates for exceeding a 15% decrease in annual precipitation (right), for the RCP6.0, based on the precalculated TCM runs.

*Supplementary Tables*

Table S1. Parameter values of the five box carbon cycle model.

| $\alpha_1$ | 0 | $\gamma_1$ | 0.13 |
|---|---|---|---|
| $\alpha_2$ | $1 - e^{-1/363}$ | $\gamma_2$ | 0.2 |
| $\alpha_3$ | $1 - e^{-1/74}$ | $\gamma_3$ | 0.32 |
| $\alpha_4$ | $1 - e^{-1/17}$ | $\gamma_4$ | 0.25 |
| $\alpha_5$ | $1 - e^{-1/2}$ | $\gamma_5$ | 0.1 |
| $\beta$ | 0.00045 | | |

Table S2. List of climate models included in the CMIP5 experiment that are emulated in AIRCC-Clim.

| ACCESS1-0 | CSIRO-Mk3-6-0 | GISS-E2-R_p1 | MIROC-ESM |
|---|---|---|---|
| ACCESS1-3 | EC-EARTH | GISS-E2-R_p2 | MIROC-ESM-CHEM |
| BNU-ESM | FGOALS-g2 | GISS-E2-R_p3 | MPI-ESM-LR |
| CanESM2 | FIO-ESM | HadGEM2-AO | MPI-ESM-MR |
| CCSM4 | GFDL-CM3 | HadGEM2-CC | MRI-CGCM3 |
| CESM1-BGC | GFDL-ESM2G | HadGEM2-ES | NorESM1-M |
| CESM1-CAM5 | GFDL-ESM2M | IPSL-CM5A-LR | NorESM1-ME |
| CMCC-CM | GISS-E2-H_p1 | IPSL-CM5A-MR | |
| CMCC-CMS | GISS-E2-H_p2 | IPSL-CM5B-LR | |
| CNRM-CM5 | GISS-E2-H_p3 | MIROC5 | |